\documentclass{ieeeaccess}
\usepackage{cite}
\usepackage{amsmath,amssymb,amsfonts}
\usepackage{algorithmic}
\usepackage{graphicx}
\usepackage{textcomp}
\usepackage{caption}
\usepackage{multirow}
\usepackage{enumitem}
\def\BibTeX{{\rm B\kern-.05em{\sc i\kern-.025em b}\kern-.08em
    T\kern-.1667em\lower.7ex\hbox{E}\kern-.125emX}}
\begin{document}

\begin{center}
\textbf{This article has been accepted for publication in IEEE Access. This is the author's version which has not been fully edited and content may change prior to final publication.}
\end{center}

\history{Date of publication xxxx 00, 0000, date of current version xxxx 00, 0000.}
\doi{10.1109/ACCESS.2017.DOI}

\title{Explainable Artificial Intelligence (XAI) for Malware Analysis: A Survey of Techniques, Applications, and Open Challenges}

\author{\uppercase{Harikha~Manthena}\authorrefmark{1*},
\uppercase{Shaghayegh~Shajarian}\authorrefmark{1*},
\uppercase{Jeffrey~Kimmell}\authorrefmark{2},
\uppercase{Mahmoud~Abdelsalam}\authorrefmark{1},
\uppercase{Sajad~Khorsandroo}\authorrefmark{1},
and \uppercase{Maanak~Gupta}\authorrefmark{2}, \IEEEmembership{Senior Member, IEEE}}

\address[1]{North Carolina Agricultural and Technical State University, Greensboro, North Carolina, USA}
\address[2]{Tennessee Tech University, Cookeville, Tennessee, USA}
\address[*]{These authors contributed equally to this work.}

\markboth
{Manthena \headeretal: Preparation of Papers for IEEE ACCESS}
{Manthena \headeretal: Preparation of Papers for IEEE ACCESS}

\corresp{Corresponding author: Shaghayegh Shajarian (e-mail: sshajarian@aggies.ncat.edu).}

\tfootnote{This work is supported by Google and NSF grants 2416992, 2230610, 2113945, and 2200538 at North Carolina A\&T State University and 2416990 and 2230609 at Tennessee Tech University.}

\begin{abstract}
Machine learning (ML) has rapidly advanced in recent years, revolutionizing fields such as finance, medicine, and cybersecurity. In malware detection, ML-based approaches have demonstrated high accuracy; however, their lack of transparency poses a significant challenge. Traditional black-box models often fail to provide interpretable justifications for their predictions, limiting their adoption in security-critical environments where understanding the reasoning behind a detection is essential for threat mitigation and response. Explainable AI (XAI) addresses this gap by enhancing model interpretability while maintaining strong detection capabilities. This survey presents a comprehensive review of state-of-the-art ML techniques for malware analysis, with a specific focus on explainability methods and research mainly from 2018 to 2024. We examine existing XAI frameworks, their application in malware classification and detection, and the challenges associated with making malware detection models more interpretable. Additionally, we explore recent advancements and highlight open research challenges in the field of explainable malware analysis. By providing a structured overview of XAI-driven malware detection approaches, this survey serves as a valuable resource for researchers and practitioners seeking to bridge the gap between ML performance and explainability in cybersecurity.
\end{abstract}

\begin{keywords}
Explainable Malware Analysis, Interpretable Malware Analysis, Explainable AI, AI for Security, Malware Detection, and Malware Classification.
\end{keywords}                                                                                                                                                                                                                                                                                    
\titlepgskip=-15pt

\maketitle

\section{Introduction}

\PARstart{I}{n} today’s digital landscape, malware remains a formidable threat, causing billions in financial losses and disrupting critical services worldwide. The increasing sophistication of attacks, particularly zero-day malware, has rendered traditional detection and analysis methods increasingly ineffective. As a result, there is a need for advanced, automated malware detection solutions that can adapt to evolving threats.

Machine Learning (ML) and Deep Learning (DL) have emerged as powerful tools for malware detection, demonstrating the ability to identify both known and zero-day threats. However, to achieve high accuracy, these models often grow complex and opaque, making it difficult to understand how predictions are made. DL-based models, in particular, are frequently described as black boxes, as their decision-making processes remain largely inscrutable to users and security professionals alike \cite{roscher2020explainable}. This lack of interpretability poses a significant challenge in cybersecurity, where understanding why a detection occurred is just as important as the detection itself for ensuring reliability, fairness, and error analysis.

To address this issue, Explainable AI (XAI) has gained increasing attention. XAI aims to bridge the gap between accuracy and explainability by providing transparent, comprehensible explanations for model predictions. By making malware detection systems more explainable, XAI enhances trust, facilitates threat analysis, and enables security professionals to make informed decisions based on model outputs \cite{saranya2023systematic}.  

Several surveys \cite{das2020opportunities, xu2019explainable, islam2021explainable, weber2023beyond, chamola2023review} have reviewed XAI, covering research areas, methods, and opportunities with mathematical and visual explanations. Danilevsky et al. \cite{danilevsky2020survey} focus on XAI in NLP, while Mohseni et al. \cite{mohseni2021multidisciplinary} propose an evaluation framework. Li et al. \cite{li2020survey} explore knowledge-driven and data-driven methods, and Confalonieri et al. \cite{confalonieri2021historical} discuss XAI’s evolution in expert systems. Speith \cite{speith2022review} provides a taxonomy of XAI approaches and challenges.

Saeed and Omlin \cite{saeed2023explainable} review XAI challenges and research directions, while Milani et al. \cite{milani2024explainable} survey explainable reinforcement learning. Nasser and Nasser \cite{nasser2023toward} examine hardware-assisted ML for malware detection. Charmet et al. \cite{charmet2022explainable} explore XAI’s role in cybersecurity but do not specifically address explainable ML in malware analysis.

Despite growing interest in XAI (2018–2024), no survey exclusively focuses on malware analysis. Existing literature lacks a clear distinction between ‘interpretable’ and ‘explainable’ AI, except for Lin and Chang \cite{lin2021towards}, who categorize interpretable malware detectors. Our survey fills this gap by covering both interpretable and explainable methods in malware classification and detection, introducing a taxonomy for explainability approaches, and summarizing recent advancements. This work aims to provide a comprehensive view of explainable malware analysis, its methodologies, and open research challenges.

Hence, this paper contributes significantly to the field of XAI with a particular focus on malware analysis. The key contributions are as follows:
 
\begin{itemize}
    \item Our work presents an extensive survey covering various XAI models and techniques used across multiple disciplines. This contribution offers a broad view of XAI and showcases its applications and relevance in different areas.
    
    \item We provide an in-depth overview of ML-based approaches in malware detection to understand the intersection of ML and malware detection, which fills a gap in the current literature.
    
    \item Our research identifies key limitations and challenges in the area of explainable malware detection. We specifically point out the predominant focus on Android-based malware in existing research, suggesting a need for a more diversified approach in future studies.
    
    \item The paper also explores potential avenues for future research in XAI applied to malware detection. We emphasize less-explored areas, such as malware detection for Windows, PDF, Linux, and hardware, thereby encouraging further investigation and development in these domains.
\end{itemize}

Our survey method involves a detailed search across various academic databases and platforms, including Google Scholar, IEEE Xplore, Science Direct, ResearchGate, arXiv, ACM, and Springer. We focus our search using a series of targeted keyword parameters. These keywords were chosen to cover a wide range of pertinent subjects. They included terms such as "explainable machine learning," "explainable artificial intelligence," "XAI," explainable malware," "explainable malware analysis," "explainable malware detection," and "explainable AI on malware detection." In addition, we also used keywords like "interpretable machine learning," "interpretable artificial intelligence," and "interpretable malware analysis." This approach allowed for an extensive and systematic review of the literature in the domains of explainable and interpretable ML and AI, with a particular emphasis on malware analysis and detection.

The structure of the remainder of this paper is outlined as follows: Section \ref{sec:maldetc} offers an in-depth exploration of file classification and online malware detection methods. Section \ref{sec:expmachinelearning} discusses ML-based models and the explainable techniques. Section \ref{sec:expmalanalysis} is dedicated to the studies on approaches and techniques in explainable malware classification and detection. This is followed by Section \ref{sec:futuredirections}, which addresses the open challenges and future prospects in this area. Finally, The paper concludes with a conclusion, which provides a summary of our work.
\section{Malware Detection Approaches}
\label{sec:maldetc}

Malware detection techniques are used to detect the threat posed by malware. They are generally categorized into two distinct approaches: File Classification and Online-Based Approaches. The field has seen considerable research efforts, with numerous studies and developments aimed at enhancing the efficacy and reliability of these malware detection methodologies. 

\begin{figure*}
    \centering
    \includegraphics[scale = 0.9]{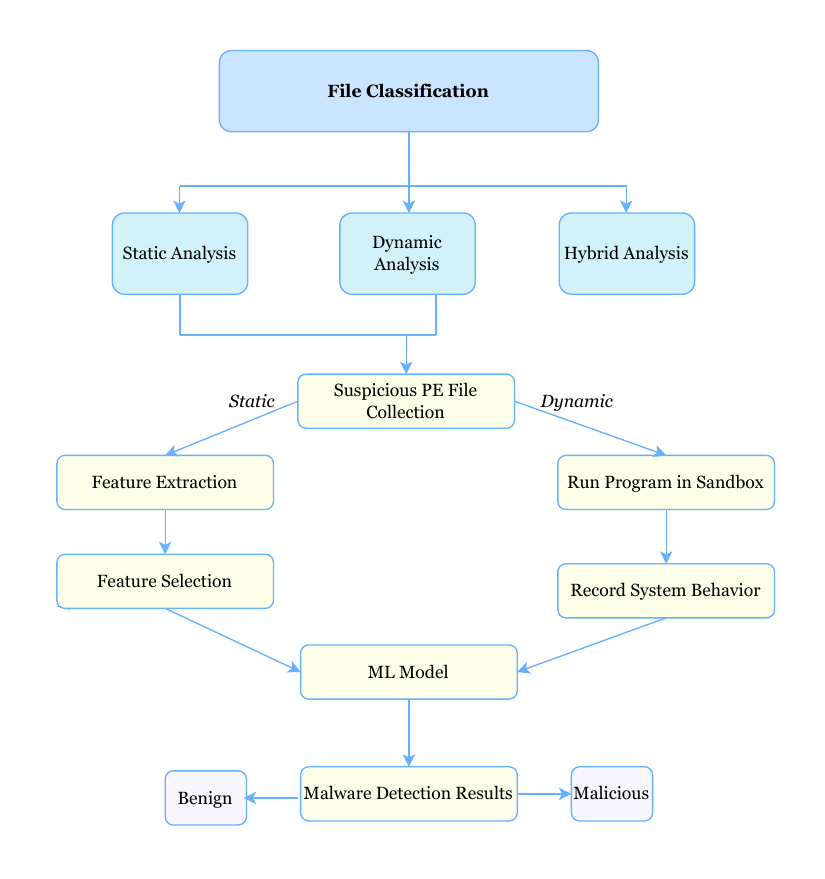}
    \caption{Machine Learning-Based File Classification Techniques}
    \label{fig:Malware Detection}
\end{figure*}

\subsection{File Classification Approach}

File classification focuses on the analysis of a file's code to determine whether it is malware. The process begins with the identification of a potentially suspicious file. To thoroughly assess its nature, file classification employs different methods, which fall into three main categories: Static analysis, Dynamic analysis, and Hybrid analysis. 

Static analysis involves examining the file's code without executing it and looking for malicious patterns. In contrast, in dynamic analysis, the file is executed in a secured environment to observe and analyze its behavior. Hybrid analysis combines these two approaches, leveraging the strengths of both static and dynamic examinations. Once a file is concluded to be non-malicious, it is generally exempt from ongoing scrutiny. These varying techniques in file classification are designed to address different aspects of malware detection. This categorization, along with the explanation of dynamic and static analysis processes, are also illustrated in Fig \ref{fig:Malware Detection}.

\subsubsection{Static Analysis}

Static analysis involves the careful examination of an executable's signature without the need to execute the code, aiming to classify the file as malware if the signature appears malicious or as benign if otherwise \cite{damodaran2017comparison}. This method has the reverse engineering of malware code and involves the detailed processing of extracted features to discern and interpret any malicious activities through a signature-based approach. In this context, a signature refers to a unique identifier for a binary file, determined by calculating its cryptographic hash. 

Multiple research, e.g. studies by Hou et al.\cite{hou2017deep} and Kim and Lee \cite{kim2021research}, have been dedicated to enhancing static malware detection, with a particular focus on the extraction of Application programming interfaces (API) calls from Portable Executable (PE) files using techniques like stacked autoencoders. This process involves extracting vital features such as API calls, Opcode sequences, and N-Grams from potentially suspicious files, as illustrated in Fig \ref{fig:Malware Detection}, which are then employed to train ML algorithms for more accurate and efficient malware detection.

For instance, the work of Shankarapani et al. \cite{shankarapani2011malware} has been using API and Opcode sequences to effectively identify segments of code that closely resemble known malware patterns. However, it is important to recognize that static analysis, while valuable, is not without its limitations. One significant challenge is its inability to detect malware that is actively running within a system or to identify completely new malware variants that have not been cataloged.

\subsubsection{Dynamic Analysis}

It involves executing malware within a secure virtual environment, such as a cuckoo sandbox, to study its behavior meticulously \cite{damodaran2017comparison} \cite{kakisim2018analysis}. This method is particularly effective in addressing zero-day malware threats. The dynamic analysis process, as depicted in Fig \ref{fig:Malware Detection}, starts with executing a suspicious PE file in a sandbox environment, ensuring isolation from external systems. This controlled execution allows for the collection of essential data, including memory features, system calls, and function calls. Subsequently, these collected data are preprocessed and used to train various ML-based algorithms, which can enhance the malware detection model.

Unlike static analysis, dynamic malware analysis requires the execution of code in a time-restricted, closed environment, which can be resource-intensive. Research endeavors,  such as studies by Firdausi et al. \cite{firdausi2010analysis} and Luckett et al. \cite{luckett2016neural}, have utilized system calls as key features for training traditional ML models like k-Nearest Neighbor (k-NN), Decision Tree, Support Vector Machine (SVM), and Naive Bayes. Furthermore, studies by Pirscoveanu et al. \cite{pirscoveanu2015analysis}, and Tobiyama et al. \cite{tobiyama2016malware} have focused on the extraction of features from API calls, evaluating the effectiveness of ML algorithms, including Random Forest, k-NN, and Convolutional Neural Networks (CNN) in dynamic malware analysis. These approaches highlight the dynamic method's capacity for dealing with complex malware detection challenges, although it requires significant time and resource allocation.

\subsubsection{Hybrid Analysis}

A methodology integrating static and dynamic techniques, hybrid analysis, is another malware detection technique \cite{tzermias2011combining}. This concept has been explored in various studies. For example, Santoso et al. \cite{santoso2019malware} utilized a combination of Artificial Neural Networks (ANN) and CNN for malware detection. Focusing on Android malware, Zhu et al.\cite{zhu2021hybrid} proposed an innovative framework using the Merged Sparse Autoencoder (MSAE), which is an unsupervised learning algorithm demonstrating its effectiveness.

Adding to this, Tong and Yan \cite{tong2017hybrid} developed a method that combines static and dynamic analysis for mobile malware detection. This method compares system call patterns of benign and malicious applications with the dynamic analysis applied to unknown applications. Subsequent offline comparison of these pattern sets further validates the unknown application's nature. Their results show the advantages of the hybrid approach over methods relying solely on static or dynamic analysis. Altaher and Barukab \cite{altaher2017intelligent} also proposed a hybrid methodology for Android malware detection that leverages API calls and application permissions, further substantiating the potential of hybrid techniques in this field.

\subsection{Online Malware Detection}

Online malware detection stands out as a distinct approach in the cybersecurity domain. Unlike static, dynamic, or hybrid methods from file classification that analyze specific malware samples, online detection monitors the entire system in real-time, which enables the capture of malware at any moment, regardless of its activity level. This technique focuses on the behavior of the entire machine rather than individual malware behaviors.

Key contributions in this area include the work of Watson et al. \cite{watson2015malware}, who developed a system using performance metrics to build SVM, achieving a 90\% accuracy rate. Azmandian et al. \cite{azmandian2011virtual} proposed intrusion-based detection techniques, while Abdelsalam et al. \cite{abdelsalam2017clustering} introduced a sequential k-means clustering algorithm for anomaly detection, specifically designed for a standard 3-tier architecture on an OpenStack Testbed. Their approach leverages virtual machine systems and resource utilization features but shows limitations in detecting low-resource-utilization malware.

Further research was done by McDole et al. \cite{mcdole2020analyzing}, who examined various CNN models to determine their suitability for malware detection in cloud Infrastructure as a Service (IaaS). Their subsequent study \cite{mcdole2021deep} compared the process-level performance metrics of different deep learning models in the context of online malware detection in cloud IaaS environments. 

Similarly, Kimmel et al. \cite{kimmell2021analyzing, kimmel2021recurrent} presented a comprehensive analysis of the effectiveness of several ML models for online malware detection, focusing on system features describing processes in a virtual machine. They emphasized the use of CNNs, which are known for their simplicity and effective representation in 2D format. Abdelsalam et al. \cite{abdelsalam2018malware} extended this concept by employing a 3-dimensional CNN to enhance classifier accuracy and specifically target low-profile malware, achieving an accuracy rate of 90\%.

\section{Explainability in Machine Learning}
\label{sec:expmachinelearning}
\begin{figure*}
    \centering
    \includegraphics[scale=0.4]{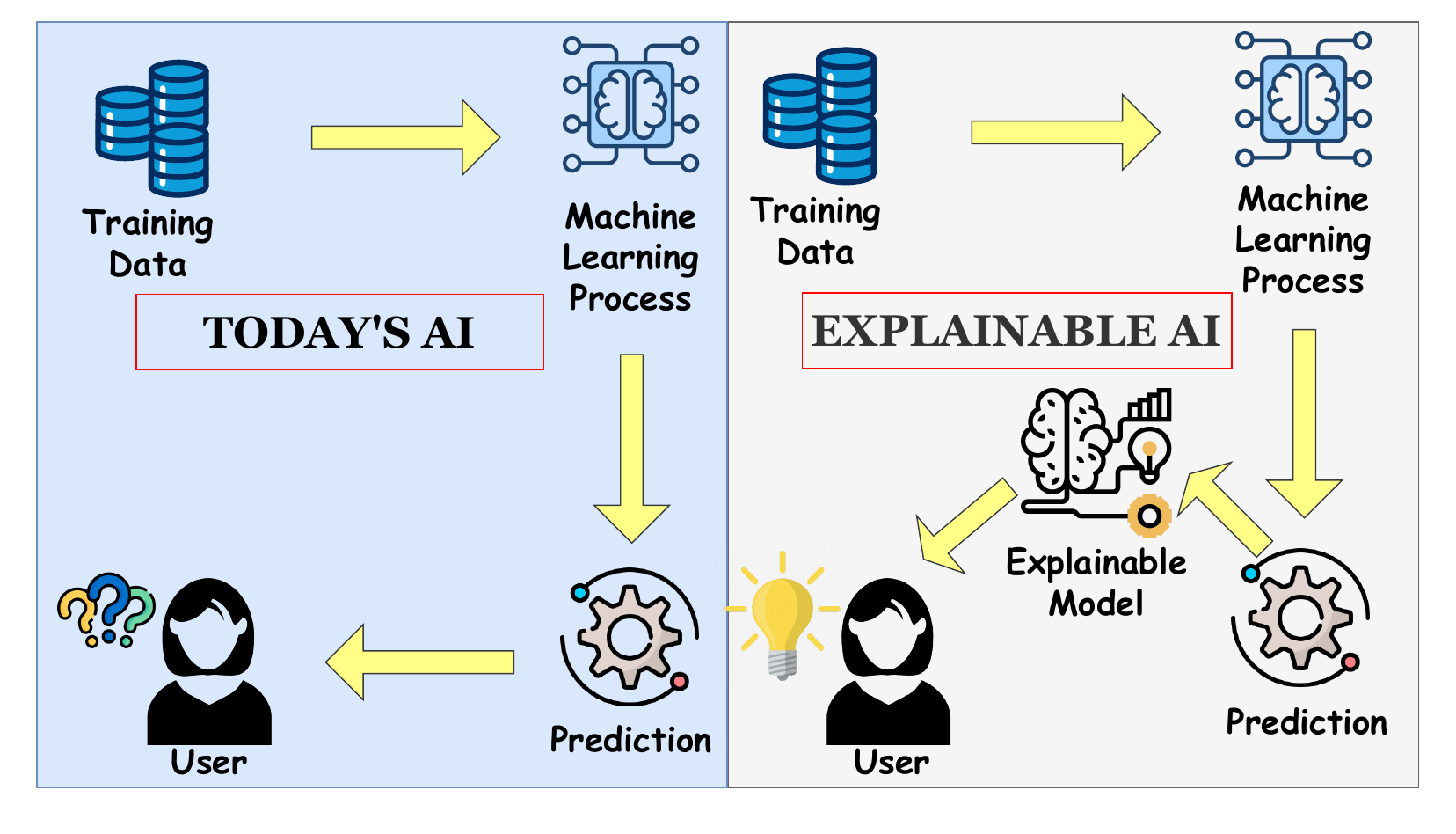}
    \caption{Explainable Machine Learning Concept}
    \label{fig:XAI_Concept}
\end{figure*}

DL-based models enable machines to develop complex hierarchical data patterns, which play a key role in tasks like classification or detection. These black-box models, by layering and integrating various levels of data representation, can enhance the predictive power of systems. However, this increased complexity often obscures the internal decision-making process, which can lead to questions about their decision logic \cite{linardatos2021explainable,arrieta2020explainable}. 

In contrast, white-box models offer a more transparent approach. They are designed to be easily interpretable, which allows users to understand how input data is transformed into predictions or decisions. This transparency is particularly valuable in fields where understanding the reasoning behind a decision is as important as the decision itself \cite{hassija2023interpreting, dwivedi2023explainable}.

For instance, in the context of cancer diagnosis, medical professionals often rely on predictive models. While these models are useful tools, there is always a possibility of incorrect predictions. Therefore, both practitioners and patients have to trust these models, which becomes possible in the situation that they understand the underlying reasons for their predictions. 

This is where the concept of XAI comes into the picture. As depicted in Fig \ref{fig:XAI_Concept}, today's AI systems typically involve training the data, undergoing the machine learning process, and providing prediction for end-users. In contrast, XAI goes a step further. It can deliver high-accuracy predictions and provides clear, justifiable explanations for these outcomes. With higher interpretability, the reasons behind AI predictions become more comprehensible to humans, which boosts the trustworthiness and reliability of the model's predictions.

In the comprehensive study, Blanco-Justicia and Domingo-Ferrer. \cite {blanco2019machine} discussed the seven characteristics that define XAI for enhancing transparency and efficacy in AI systems as follows.

\textbf{Accuracy.} This aspect evaluates how well an XAI model predicts outcomes for new, unseen data. Predictions made by these models must have a high level of accuracy.

\textbf{Fidelity.} It is about the closeness of the explanation to the model’s prediction. An explanation is regarded as highly accurate when it meets the high fidelity and high accuracy of the black-box model.

\textbf{Consistency.}  This characteristic describes how equally explanations are applied to a model that is trained on the same dataset.

\textbf{Stability.} It examines whether the stability is reflected in the explanation model, which means that similar instances should produce similar explanations.

\textbf{Degree of Importance.} This attribute indicates how well the explanation reflects the significance of various features within the model, which is essential for understanding the weight of different aspects in the model's decision-making process.

\textbf{Novelty.} Closely related to stability, novelty assesses the ability of the explanation mechanism to accurately represent data instances that are significantly different from those in the training set.

\textbf{Representativeness.} This factor has a significant effect on explainability, emphasizing the need for explanations to be relevant and applicable in a diverse range of decision-making scenarios, thereby ensuring their utility across various applications. 

\begin{figure*}
    \centering
    \includegraphics[scale = 0.5]{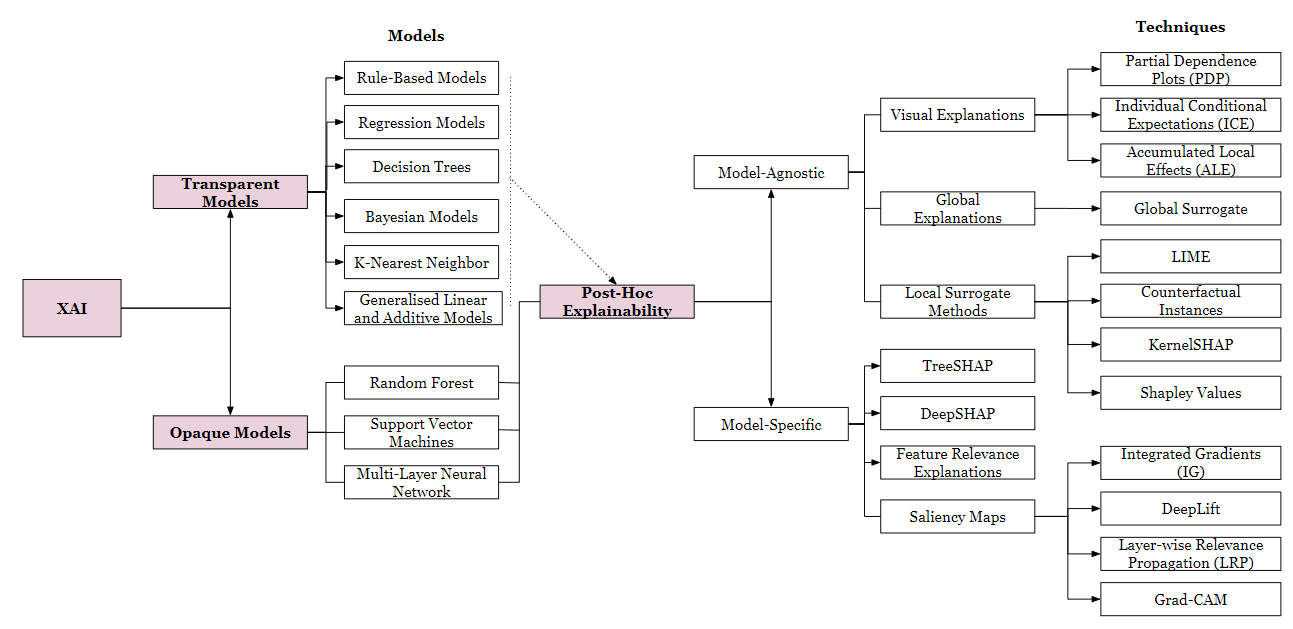}  
    \caption{Taxonomy for Explainable Machine Learning Techniques Inspired by \cite{belle2021principles}, \cite{chou2022counterfactuals} and \cite{arrieta2020explainable}.}
    \label{fig:XAI_Taxonomy}
\end{figure*}

Hence, in the context of XAI, it is important to understand the general classification of ML-based models, which is illustrated in Fig \ref{fig:XAI_Taxonomy}.  This figure presents a comprehensive taxonomy of ML models and XAI techniques and provides a clear framework for understanding this field.

As depicted in Fig \ref{fig:XAI_Taxonomy}, ML models can be broadly classified into Transparent and Opaque models.  Transparent models are inherently explainable. These models are straightforward enough that they do not require additional post-hoc explainability techniques, i.e., techniques provide explanations only after the training process has finished. However, as indicated by the dashed arrow in Fig \ref{fig:XAI_Taxonomy}, when these models become more complex, post-hoc explainability may still be needed to improve clarity and ensure human interpretability.

On the other hand, Opaque models, often referred to as black-box models, are characterized by their high accuracy yet present challenges in interpretation. Due to their complexity, they require the use of post-hoc explainability methods. The goal of post-hoc explainability is to make the outcomes of ML-based models more transparent, understandable, and trustworthy to humans. 

Post-hoc explainability can be further divided into two types: model-agnostic and model-specific methods. Model-agnostic methods have a variety of explainability techniques and are versatile enough to be applied to any black-box model. In contrast, model-specific methods are applicable only to certain types of models and limit their utility to specific cases.

The subsequent section of this paper will outline the various ML models and post-hoc explainability techniques, providing a comprehensive summary of different research challenges encountered in this evolving field. 

\subsection{Transparent Machine-Learning Models}

Transparent models are distinguished by their inherent ability to be self-explanatory. They can be interpreted directly, enabling users to comprehend their decision-making processes. This category of models, from Rule-based Learners and Regression Models to Decision Trees, Bayesian Models, k-NN algorithms, and the Generalized Additive Model (GAM), are unified by their transparent nature.  

\subsubsection{Rule-Based Models}

These models are characterized by developing rules to represent and interpret the data they are designed to learn from. At the core of these models is the IF-THEN statement, a basic but powerful structure that forms the foundation of these rules. The IF part represents the condition, while the THEN part denotes the prediction. These predictions can arise from a single rule or a synergy of multiple rules. Soares et al. \cite{soares2020explaining} apply this concept to explain DL-based models. They propose a method where a deep reinforcement learning model is approximated through a series of IF-THEN rules, effectively enhancing the model's interpretability. 

The clarity of rule-based models makes them highly interpretable and understandable. Their straightforward structure eliminates the need for post-hoc analysis. They are also used to clarify the predictions of more intricate models by generating and applying rules to link sophisticated ML-based techniques with approachable interpretability.

\subsubsection{Regression Models}

Linear and logistic regression models stand as two important regression models. The weight of the coefficient of linear regression is easy to quantify and interpret, which is why it is used in various fields to explain the predictions. On the other hand, logistic regression strength lies in its ability to provide probabilities alongside classifications, which offers a nuanced view of outcomes.

Lundberg's research \cite{lundberg2019explainable} further enhances this model's utility by integrating logistic regression with gradient-boosted trees for predicting synthetic labels and augmenting the explainability of tree-based models. Despite their transparency, these regression models often require additional post-hoc explainability tools, like visual aids, to make their predictions accessible to those not well-versed in statistical methodologies.

\subsubsection{Decision Trees}

Decision trees offer transparent models that enable domain experts to understand how they work. Furthermore, the exploration of these trees can lead to the discovery of new relationships and insights. Blanco-Justicia and Domingo-Ferrer \cite{blanco2019machine} leverage decision trees as surrogate models to elucidate black-box models, constructing these trees from segmented portions of the training dataset.

This approach assumes that the person responsible for providing explanations has access to the training data and the black-box model. Nevertheless, decision trees encounter scalability issues with large datasets in real-world applications, which diminish their explainability as the tree complexity increases. This complexity requires the adoption of post-hoc explainability methods to maintain clarity. 

\subsubsection{Bayesian Models}

Bayesian models excel in providing a high degree of interpretability and explainability, offering insights into the statistical interplay between variables. This capability makes them particularly useful for applications where clear, comprehensible explanations are essential, such as demonstrating the correlation between diseases and their symptoms. 

Hence, in the realm of medical research, the application of Bayesian methods has been notably effective.  For instance, Arrieta et al. \cite {arrieta2020explainable} demonstrate the utility of Bayesian approaches in healthcare analytics, highlighting their potential to the complex relationships within medical data. Similarly, the Naive Bayes classifier, as discussed by Rana et al. \cite {rana2018evaluating},  serves as a robust algorithm for predictive modeling. This classifier efficiently tackles both binary and multi-classification problems by calculating the probabilities of individual elements, subsequently employing Bayes' theorem to identify the most probable outcome.

\subsubsection{k-Nearest Neighbour}

The k-NN algorithm operates on a simple yet effective principle: for classification, it determines a test sample's class based on the majority vote from its nearest neighbors, and for regression, it computes the average outcome of these neighbors.  The interpretability of k-NN is significantly influenced by the chosen features, the distance metric utilized, and the number of neighbors. While models with extensive features may obscure interpretability, a k-NN model characterized by a concise, well-selected feature set remains one of the main models for interpretable results. In a study, Aslam et al. \cite{aslam2022explainable} showcase the application of various supervised ML-based models, including k-NN, utilizing XAI techniques.

\subsubsection{Generalized Linear and Additive Models}

The Generalized Additive Model (GAM) represents an advancement in statistical modeling, combining the benefits of linearity and interpretability. This model assigns values to variables by integrating numerous undefined functions specific to regression models and enhancing accuracy without compromising interpretability. One of the unique features of GAM is its ability to allow users to evaluate the significance of each variable by examining its impact on the predicted outcome. GAM models exhibit algorithmic transparency and are regarded as simulatable due to their minimal dimensionality issues. 

To have an optimal balance between accuracy and explainability, Yang et al. \cite {yang2021gami} introduce GAMI-Net (Generalized Additive Models with Structured Interactions), a neural network characterized by its intrinsic explainability. This model has been benchmarked against several standard models, including GLM, showcasing its robustness.

Despite the inherent transparency and explainability of these models, ongoing research explores undirected graphical models to further enhance their trustworthiness. Transparency alone may not always guarantee straightforward explainability, as increasing model complexity can reduce interpretability. This necessitates the development of post-hoc explanations to maintain clarity and reliability. As mentioned above and illustrated in Figure \ref{fig:XAI_Taxonomy}, the dashed arrow represents cases where even transparent models may require post-hoc explanations in complex scenarios. This highlights the dynamic nature of explainability, where certain conditions still necessitate additional interpretability techniques to ensure model reliability and trust.

\subsection{Opaque ML-based Models}

We explored models characterized by their transparency,  highlighting that their interpretability does not guarantee enhanced performance. This section shifts focus to examine complex models that stand out for their high accuracy. However, these models require post-hoc explanations to unlock an understanding of their internal processes.

\subsubsection{Random Forest}

Random Forests (RFs) consist of multiple decision trees, each dividing the input space into smaller segments and averaging outcomes. As problem complexity increases, more trees are needed, improving accuracy but reducing explainability. RFs were designed to mitigate overfitting in single decision trees by averaging predictions across multiple trees, reducing variance. Each tree is trained on a unique data subset, ensuring diverse insights. 

However, the model's complexity necessitates post-hoc explainability techniques. Zhao et al. \cite{zhao2018iforest} introduce a visual analytic system to enhance interpretability, offering a comprehensive approach to understanding RF predictions.

\subsubsection{Support Vector Machine}

SVM constructs a hyperplane or a set of hyperplanes within a high or infinite-dimensional space, serving purposes across classification, regression, outlier detection, and even clustering tasks. A hyperplane achieves optimal separation when it maximizes the distance to the nearest point of the training dataset, as a larger margin correlates with a lower generalization error of the classifier. Owing to their remarkable predictive and generalization capabilities, SVMs are among the most widely utilized ML models. However, due to their complex dimensionality, they are often regarded as opaque, making their decision-making process less transparent. 

Based on this, Vieira and Digiampietri \cite{vieira2020study} explore the use of decision trees to derive rules from SVMs, which provides explanations for the classifications made by SVM classifiers and enhances their interpretability.

\subsubsection{Multi-Layer Neural Network}

These models are computationally intensive but provide unparalleled performance across a wide range of applications. Neural networks are inherently considered black-box models due to their complex internal mechanisms. In the study by Sharma et al. \cite{sharma2020building}, the focus is on utilizing a multi-layer perceptron neural network for the risk prediction of default loans, with the explanation of model decisions facilitated through a sensitivity analysis technique.

\subsection{Model-Agnostic Techniques for Post-Hoc Explainability}

Post-hoc explainability methods play a crucial role in interpreting complex ML-based models, especially in high-stakes applications where decision transparency is required. One way to classify post-hoc explainability methods is based on their dependency on the model structure. In this classification, we identify two groups: model-agnostic methods, which can be applied to any model, and model-specific methods, which are tailored to particular model architectures.

Model-agnostic techniques can be applied across various ML architectures, offering greater flexibility. However, this flexibility often comes at the cost of precision in explanations, as these methods approximate model behavior rather than providing direct interpretability. Given the increasing reliance on ML in security-sensitive domains such as malware detection, adopting robust and interpretable post-hoc techniques is essential to ensure trust and accountability. The domain of model agnostic interpretability is divided into three main categories: Global Explanation, Local Explanation, and Visual Explanation.

\subsubsection{Global Explanation}

Global explainability provides insights into how a model makes decisions across all instances, rather than focusing on individual predictions. This is essential for understanding which features are most influential and how they interact at a systemic level.

A widely used approach to achieving global explainability is the use of surrogate models, which approximate the decision-making process of a complex black-box model by training a more interpretable alternative (e.g., decision trees, linear models) on the same dataset. These models enable researchers to analyze feature contributions in a transparent manner, facilitating model interpretation.

However, surrogate models introduce trade-offs. Since they approximate rather than replicate the black-box model’s decision boundaries, they may introduce inaccuracies, particularly in the presence of nonlinear relationships or complex feature interactions. Despite these limitations, they remain a practical tool for obtaining high-level insights into opaque models.

In cybersecurity and malware detection, global explanations help identify critical risk factors in security assessments. For instance, a surrogate model trained on malware classification outputs can reveal whether API call sequences, file metadata, or network behaviors are the most significant predictors. This understanding allows security analysts to refine detection rules, improve feature selection, and enhance model robustness.

Surrogate models serve as interpretable stand-ins for complex models, offering insights into their underlying mechanisms. For example, Islam et al. \cite{islam2021explainable} demonstrated the effectiveness of this approach by using Classification and Regression Trees (CART) to approximate a random forest’s decision-making process. If a surrogate achieves comparable performance, it may reduce reliance on the original model, particularly when interpretability is prioritized. Additionally, multiple surrogate models can be developed for a single black-box system, each providing distinct perspectives on model behavior. This approach enhances transparency and facilitates the comprehension of sophisticated decision processes.

\subsubsection{Local Explanation}

Local explainability focuses on understanding why a model makes a specific prediction for a single instance rather than explaining overall model behavior. These methods help users answer questions such as, "Why was this particular file classified as malware?" or "What features contributed to this anomaly?" Unlike global explanations, which provide an overview of feature importance across an entire dataset, local explanations offer insights into decision-making at an individual level.

Local explanation techniques are particularly valuable in high-stakes applications like cybersecurity, where understanding why a model flagged a file as malicious can assist analysts in investigating threats, identifying adversarial attacks, or refining detection rules. Additionally, local explanations play a critical role in bias detection and fairness assessments, ensuring that models do not make decisions based on unintended or discriminatory features.

Several widely used model-agnostic techniques exist for local explainability, including LIME (Local Interpretable Model-Agnostic Explanations), KernelSHAP (Shapley Additive Explanations), Shapley values, counterfactual explanations, and Logic Explained Networks (LENs). The following sections discuss these methods in more detail, highlighting their strengths, limitations, and practical applications.

\textbf{LIME.} For the first time, Ribeiro et al. \cite {ribeiro2016should} introduce a novel local explainability technique known as LIME. This method operates as a local surrogate model, generating interpretable predictions by approximating how the model behaves in the vicinity of a given prediction. It is designed to be model-agnostic, which makes it versatile across different ML-based models.

To evaluate the effectiveness of the surrogate model, LIME employs a local fidelity measure. This metric assesses the extent to which LIME's approximations reflect the true behavior and accuracy of the underlying black-box model. However, it is important to note that LIME is not equipped to offer insights into the global operations of a model. Furthermore, if the local fidelity measure indicates poor accuracy, the reliability of LIME's interpretability may be compromised.

In a practical application of LIME, Magesh et al. \cite{magesh2020explainable} utilize this technique to interpret the predictions of a CNN model designed for the early detection of Parkinson's disease. This study demonstrates the potential of LIME to provide valuable insights into real-world scenarios.

\textbf{KernalSHAP.} Among the various local interpretability methods developed, a significant challenge lies in determining the most suitable method for specific scenarios. To address this, Lundberg and Lee \cite{lundberg2017unified} propose Shapley Additive exPlanations (SHAP), a concept derived from game theory that evaluates the importance of each feature in contributing to a particular prediction. The SHAP framework establishes a new class of additive feature importance measures characterized by a unique solution that exhibits desirable attributes.

KernelSHAP, as part of the SHAP family, is model-agnostic, allowing its application across diverse ML models. The computation of exact SHAP values via KernelSHAP can be exponentially time-consuming, which highlights its computational demands. Despite this, its capability to adapt to any ML model shows its broad utility. The SHAP framework also includes tailored variants such as TreeSHAP and DeepSHAP, designed specifically for tree-based and deep learning models, respectively. These variants can optimize the efficiency and relevance of SHAP analysis in targeted model types.

\textbf{Shapley Values.} They originate from coalitional game theory, framing each feature value of an instance as a "player" and the prediction outcome as the "payout." This approach assigns a quantifiable contribution to each feature, which can demonstrate how significantly each one influences the final prediction. Shapley values are distinguished by key principles such as consistency and local accuracy. These principles ensure that the allocation of importance to features is both fair and interpretable, accurately reflecting each feature's contribution to the outcome.

\textbf{Counterfactual Explanations.} Counterfactual explanations provide a compelling approach for local interpretation. This method stands out for its simplicity in implementation, as it does not need access to the underlying data or model. Counterfactual explanations focus on identifying which features would need alteration to achieve a specific desired outcome, thereby elucidating the reasoning behind model predictions. These explanations are particularly user-friendly because they illustrate how minimal changes in features can influence predictions. Nonetheless, one limitation of this method is its difficulty in accommodating categorical data across different levels. Related to the Counterfactual explanation, Molnar \cite{molnar2022} discusses their application in models generating continuous predictions that showcase their utility in providing clear and actionable insights. In malware classification, counterfactual explanations can highlight the minimal changes required to alter a model’s decision, offering valuable insights for both threat analysis and adversarial defense strategies. This approach is particularly useful in detecting adversarial attacks, as it helps identify which modifications in malware features could evade detection, thereby strengthening model robustness.

\textbf{LENs.} They enhance the interpretability of neural networks by utilizing human-understandable predicates as inputs and translating predictions into First-Order Logic (FOL) explanations. These networks are highly adaptable and can function effectively in both supervised and unsupervised learning contexts. LENs can serve as direct classifiers, providing explanations for their predictions, or they can work alongside black-box classifiers to make their decisions interpretable. The learning process for LENs involves associating specific input features with output classes in supervised scenarios and generating logic rules that explain the conditions for predictions. In unsupervised learning, LENs identify patterns and relationships within the data, clustering similar data points and generating explanations that describe these clusters. Additionally, LENs can mimic the outputs of black-box models while generating FOL explanations, which leads to elucidating the decision-making process for these complex models \cite{ciravegna2023logic}. LENs bridge the gap between symbolic reasoning and neural networks by incorporating logical constraints into the learning process, making their decisions more interpretable. Unlike traditional neural networks, which act as black boxes, LENs provide structured, rule-based explanations that enhance transparency and trust in AI-driven decision-making.

\subsubsection{Visual Explanation} This approach contains methods designed to produce visual representations of models that make them accessible and comprehensible. Techniques such as Individual Conditional Expectation (ICE), Partial Dependence Plot (PDP), and Accumulated Local Effects (ALE) serve as key tools in this visualization process. These techniques facilitate a deeper understanding of how models operate by graphically depicting the relationship between features and the model's predictions. The advantage of visual explanations lies in their ability to convey complex model dynamics in a manner that is easily graspable. This makes visualizing techniques invaluable for broadening the accessibility of model interpretations.

\textbf{Partial Dependence Plot (PDP).} It offers insights into the marginal impact of one or two features on the predicted outcome of an ML model, as highlighted by Molnar \cite{molnar2022}. This tool is important in determining whether the relationship between the target and features is linear or exhibits more complexity. For instance, in the context of a linear regression model, PDP can reveal a linear relationship and illustrate how variations in a specific feature correlate with changes in the prediction. Unlike methods that focus on the influence of features on individual predictions, PDP emphasizes the average effect of features on the model's overall behavior. However, its application is generally constrained to analyzing up to two features simultaneously, based on the assumption that the selected features are independent of others not included in the plot.

\textbf{Individual Conditional Expectation (ICE).} Within the post-hoc explainability, visual explanations, particularly those compatible with model-agnostic approaches, are notably rare. The ICE plot, introduced by Goldstein et al. \cite{goldstein2015peeking}, emerges as a visualization technique for delineating the predicted outcomes of models governed by supervised learning algorithms. Diverging from the PDP, the ICE plot underscores the dependency of predictions on a specific feature across individual instances, each represented by a unique line. This approach allows users to understand how changes in a feature impact predictions on a case-by-case basis.

ICE plots especially highlight the variability of predictions within the range of a given covariate, identifying areas of significant heterogeneity. This capability is complemented by a visual test for assessing the model that generated the data alongside a comprehensive suite of tools for exploratory analysis. By employing both simulated examples and real-world data sets, the creators of ICE plots demonstrate their utility in uncovering insights about estimated models that PDPs may not reveal, offering a more granular perspective on model behavior.

\textbf{Accumulated Local Effects (ALE).} The ALE plot provides a visual illustration of how individual features influence the predictions made by a machine learning model. It effectively showcases the dynamics between regressors (independent variables) and the dependent variable, offering insights into their relationship. Notably, ALE plots are recognized for their efficiency, being faster to generate compared to PDP.

Kramer et al. \cite{kramer2021explainable} demonstrated the application of ALE plots within the realm of real estate, employing them to discern which features significantly impact property values. This use case underscores the utility of ALE plots in practical, real-world analysis. Additionally, many researchers have adopted ALE plots as a method for visually exploring the nature of relationships between variables, assessing whether these relationships are linear or exhibit more complexity. 

\subsection{Model-Specific Techniques for Post-Hoc Explainability}

Model-specific methods of post-hoc explainability are designed to be applied exclusively to certain types of models. These techniques can also be categorized based on their scope of interpretability, which includes local, global, and visual dimensions. Local scope refers to methods that focus on explaining the prediction for an individual data point. In contrast, global scope encompasses techniques that interpret the overall behavior of the model. Meanwhile, visual scope techniques are aimed at creating visual representations that make model behaviors comprehensible.

Among the array of model-specific approaches, TreeShap and DeepSHAP are notable for their application to tree-based and deep learning models, respectively. Additionally, saliency maps encompass a variety of methods, such as DeepLift, layer-wise relevance propagation, Grad-CAM, and other gradient-based approaches, along with feature relevance explanations. 

\textbf{TreeSHAP and DeepSHAP.} They represent two specialized implementations of SHAP grounded in the principles of Shapley values. TreeSHAP is tailored for tree-based models, offering a more efficient computation of exact SHAP values by operating in polynomial time, in contrast to the exponential time typically required by the general SHAP approach. In an illustrative application, Athanasiou et al. \cite{athanasiou2020explainable} leveraged TreeSHAP within an explainable risk prediction model for cardiovascular disease, utilizing this technique to furnish personalized explanations of the machine learning model's predictions.

Conversely, DeepSHAP is devised to work with neural networks and serves as an approximation method for calculating conditional expectations of SHAP values, utilizing selected background samples for this purpose. It represents an evolution of the DeepLIFT method, adapting it to estimate Shapley values for specific inputs across the feature space. This adaptation enables DeepSHAP to pinpoint the contribution of each feature to a given prediction within neural network models. An example of DeepSHAP's application can be found in the work by Davagdorj et al. \cite{davagdorj2021explainable}, where it was employed within a neural network framework to predict non-communicable diseases. The primary objective of this approach is to elucidate the risk factors influencing the model’s predictions, aiming to provide explanations that are both meaningful and accessible to users, focusing on specific instances from the user's perspective.

\textbf{Feature Relevance Explanations.} Feature relevance explanation techniques are important in enhancing the interpretability of tree ensembles. This category contains a variety of techniques aimed at elucidating how different features contribute to a model's predictions, including feature importance, feature extraction, and feature contribution. Central to these techniques is the concept of feature importance, which assesses the significance of feature interactions in influencing the model's outcome. Adebayo and Kagal \cite{adebayo2016iterative} introduced a methodological approach for quantifying feature importance by iteratively transforming features within the dataset. This process involves eliminating features deemed non-essential, thereby creating a refined dataset that retains only those features with significant relevance.

Subsequently, the authors developed a novel metric to calculate scores for the revised datasets based on the variations observed in model performance. This approach underscores the dynamic nature of feature interactions within predictive models, where the effect of individual features on the prediction cannot simply be aggregated to reflect the total influence.

Further advancing the understanding of feature interactions, Friedman and Popescu \cite{friedman2008predictive} introduced the H-statistic. This metric is designed to explain the extent of feature interactions by measuring the variance in predictions attributable to these interactions. The H-statistic thus serves as a valuable tool for detecting and quantifying the strength of interactions among features within a prediction model.

\textbf{Saliency Maps.} They serve as critical tools in attribution analysis by showing the pixels that significantly influence image classification decisions. These gradient-based methods, designed specifically for neural network models, can facilitate an understanding of the features most relevant to a model's output. Among such methods, Layer-wise Relevance Propagation (LRP) and DeepLIFT stand out by providing a framework to assign importance scores to different elements of a network, offering a detailed explanation of a model's decision-making process. Specifically, LRP identifies the contribution of various parts of the input data towards the final decision, a technique effectively employed by Wang et al. \cite{wang2021effective} for dynamic and explainable malware detection. By pinpointing malicious code snippets, their approach enhances the interpretability of malware classifiers.

Similarly, DeepLIFT, as explored by Shrikumar et al. \cite{shrikumar2017learning}, contrasts the activation of neurons against a reference point, leveraging the differences to ascertain the significance of each feature. This method enriches our understanding of neural network operations by clarifying how each input affects the output.

Moreover, Grad-CAM represents another notable advancement in pixel-attribution methodologies, offering a refined lens through which to view the decision-making processes of CNNs. By attributing a relevance score to each neuron in the final convolutional layer and examining the activated regions within the feature map, Grad-CAM elucidates the features deemed most crucial by the CNN. This process not only aids in interpreting the model's focus but also contributes to the model's transparency.

Integrated Gradients (IG) is another one that calculates the average gradients across a straight-line path between the baseline input and the actual input. This approach, particularly beneficial for CNN predictions, highlights the incremental impact of each feature along this path, thereby offering a comprehensive view of the factors influencing the model's predictions. Collectively, these saliency map methods show the importance of model-specific analyses in enhancing the interpretability and transparency of neural networks.

\section{Explainable Malware Classification and Detection Approaches}
\label{sec:expmalanalysis}

\begin{figure*}
    \centering
    \includegraphics[width=1\linewidth]{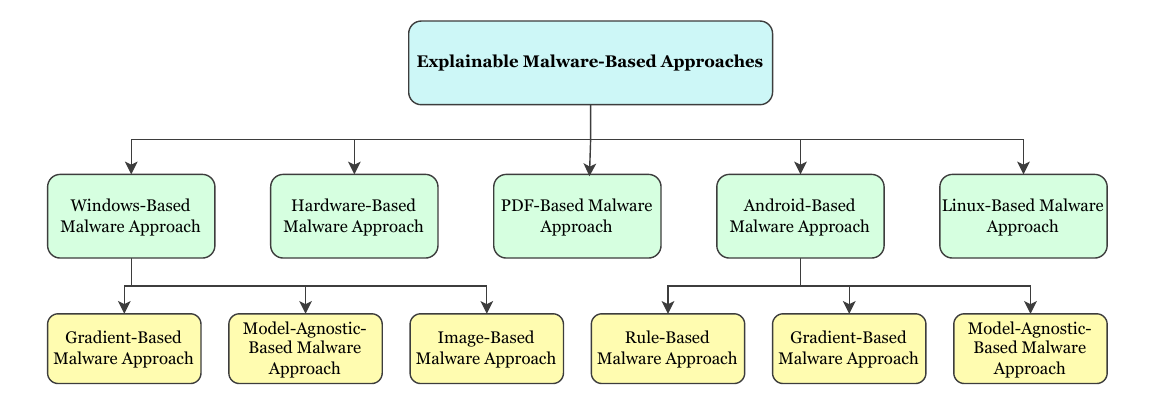}
    \caption{Explainable Malware Classification and Detection Approaches}
    \label{fig:XAI_Malware Detection Approaches}
\end{figure*}

The preceding section provided an introduction to the concept of explainability in ML, detailed various models and techniques for enhancing explainability, and reviewed relevant research in the broader field of explainable ML. Moving forward, this section will focus on the application of explainable ML in the context of malware classification and detection. These approaches are organized by the types of target systems, including Windows PE files, hardware systems, Android devices, PDF documents, and Linux files. This classification is depicted in Fig \ref{fig:XAI_Malware Detection Approaches}, which provides a clear framework for understanding how explainable ML techniques are applied across different computing platforms to address malware threats.

By categorizing existing explainable malware detection strategies by the target platform—Windows PE, Android, hardware, PDF, and Linux— we highlight how explainability methods adapt to each system’s unique execution environment and feature set. We also provide comprehensive tables (\ref{tab:Android}, \ref{tab:windows}, \ref{tab:Hardware/PDF/Linux}, \ref{tab:other approaches}) summarizing key contributions, limitations, and XAI techniques. These tables are provided for quick reference. However, we will make explanatory context in each subsection.

\subsection{Windows PE-Based Malware Approaches}

Windows is the most widely used desktop OS, making Windows PE files a significant target for malware. Below, we summarize the PE format’s structure and then highlight state-of-the-art studies based on key explainable detection methods grouped by the type of XAI approach (gradient-based, model-agnostic, and image-based). Finally, Table \ref{tab:windows} lists major works in Windows PE-based malware detection, comparing their focus, contributions, limitations, and XAI techniques.

Windows PE is a file format based on the Common Object File Format (COFF) specification and holds significant importance within the Windows operating system family. The structure of a PE file starts with a header initially used by the MS-DOS operating system. When the executable is loaded, MS-DOS runs a stub program to ensure backward compatibility. Next, the COFF header provides detailed specifications of the executable file. It is followed by an optional header, which adds flexibility and supports future enhancements to the file structure. Following this, the section header divides the executable into distinct sections. These sections comprise blocks of memory and support page swapping to address memory limitations, which leads to organizing the executable into structured segments for efficient execution.

The Windows operating system has emerged as the predominant platform on personal computers, which has increased its vulnerability to malware attacks. Despite this risk, limited research has focused on explainable malware detection methodologies specifically for Windows.

Developing ML-based models that learn discriminative features from raw inputs requires feature extraction, which is time-consuming and complex. To address this challenge, Raff et al. \cite{raff2018malware} introduce "MalConv," a novel architecture for malware detection. This architecture leverages the entire executable as input for a CNN. The MalConv architecture utilizes a methodology similar to techniques employed in speech and signal processing \cite{graves2013speech}, text understanding \cite{zhang2015text}, and image classification \cite{szegedy2015going}, where CNNs effectively extract pertinent features.

MalConv is designed as a static architecture for identifying static malware, which combines CNN activation functions with global max-pooling before progressing to fully connected layers. This approach guarantees that the ML model generates activations independent of the features' spatial locations, which leads to enhancing its ability to classify and detect malware. The discussion surrounding Windows PE malware detection includes gradient-based, model-agnostic techniques, and reliance on image representations. 

The forthcoming section will discuss an analysis of existing strategies for malware classification and detection, with Table \ref{tab:windows} explicitly addressing the application of model-agnostic, gradient-based, and image-based methodologies in the context of Windows PE-based malware research.

Table \ref{tab:windows} summarizes research on explainable Windows malware detection. Focus/Objective clarifies the primary goal, the Contribution outlines each paper’s novelty, and the XAI Technique highlights the interpretability approach used.

\begin{table*}[!htp]
    \centering
    \footnotesize
    \setlength{\tabcolsep}{0.8\tabcolsep}
    \def\arraystretch{1.5}
    \caption{Research Addressing Explainable Machine Learning in Windows Malware}
    \label{tab:windows}
    \resizebox{\textwidth}{!}{%
    \begin{tabular}{|p{1.25cm}|p{3.5cm}|p{5.5cm}|p{3cm}|p{1.5cm}|}
    \hline
    \textbf{Paper/Year} & \textbf{Focus/Objective} & \textbf{Contribution} & \textbf{Limitation} & \textbf{XAI Technique}\\
    \hline
    Bose et al. 2020~\cite{bose2020explaining} &  
    Analyzed MalConv and proposed an end-to-end gradient-based explanation framework &
    Introduced gradient analysis mapping malicious file embeddings and interpolating between correctly classified samples to clarify decision boundaries &
    Limited interpretability of Neural Network decisions &
    Gradient Analysis \\
    \hline
    Mathews 2019~\cite{mathews2019explainable} &
    Identify feature-engineering flaws and global traits distinguishing malware classes &
    Proposed an explainability framework using content-based and PE-file statistical features, with LIME for model interpretation &
    Lacks comparison with other model-agnostic XAI methods &
    LIME \\
    \hline
    Pirch et al. 2021~\cite{pirch2021tagvet} & 
    Develop an explainable CNN to predict malware tags for clustering and organization &
    Assigns a relevance score to each feature (“token”) for tagging malware, evaluated by descriptive accuracy and sparsity &
    Restricted to a behavioral-based approach &
    LRP \\
    \hline
    Marais et al. 2021~\cite{marais2021malware} & 
    Improve model interpretability to reduce false positives &
    Developed a CNN-based malware model compared with LGBM, XGBoost, and DNN, using GradCam++ to highlight critical pixels for reduced false positives &
    Missing attention mechanisms; limited to static analysis and higher computation time &
    GradCam++ \\
    \hline
    Li et al. 2021~\cite{li2021effectiveness} & 
    Introduce an interpretable feed-forward neural network (IFFNN) for malware detection &
    Proposed IFFNN with high accuracy and interpretability; tested with MNIST for qualitative assessment &
    Interpretation fidelity requires more validation &
    IFFNN \\
    \hline
    Chen et al. 2019~\cite{chen2019believe} & 
    Validate explanation fidelity in dynamic malware analysis &
    Extended LIME for vision-based interpretation in dynamic malware analysis, with two case studies showing interpretability benefits &
    Does not compare LIME with more advanced XAI methods &
    LIME \\
    \hline
    Lin and Chang 2021~\cite{lin2021towards1} & 
    Implement a deep ensemble detector for image-based malware with interpretable outputs &
    Proposed Selective Deep Ensemble Learning (SDEL) and used LIME, SHAP, and LRP to interpret image-based malware predictions &
    It needs more base classifiers and original malware files to locate semantic features &
    LIME, SHAP, and LRP \\
    \hline
    Alani et al. 2023~\cite{alani2023xmal} & 
    Detect obfuscated Windows malware from memory dumps using minimal features &
    Achieves >99\% detection using only five memory-dump features, with SHAP-based interpretability &
    Dependent on a limited feature set and dataset-specific performance &
    SHAP \\
    \hline
    Ciaramella et al. 2024~\cite{ciaramella2024explainable} & 
    Proposes a CNN-based approach for ransomware detection using image conversion &
    Uses Grad-CAM to visualize high-impact image regions, improving ransomware detection interpretability &
    Focused mainly on ransomware, with unclear generalization to other malware &
    Grad-CAM \\
    \hline
    Anthony et al. 2024~\cite{anthony2024explainable} & 
    Adapt LENs for interpretable large-scale malware detection &
    Provides high-fidelity FOL rule explanations, balancing accuracy with transparency &
    Generating/optimizing FOL rules is computationally heavy, limiting real-time application &
    LENs \\
    \hline
    Gulmez et al. 2024~\cite{gulmez2024xran} & 
    Propose XRan, an XAI-enabled CNN approach combining API, DLL, and Mutex features for ransomware detection &
    Combines API/DLL/Mutex features in a two-layer CNN, reaching 99.4\% TPR with LIME/SHAP explanations &
    High overhead for dynamic feature extraction &
    LIME, and SHAP \\
    \hline
    Aryal et al. 2024~\cite{aryal2024explainability} & 
    Use SHAP to locate and perturb critical regions of Windows PE malware for enhanced evasion &
    Demonstrates SHAP-guided perturbation boosts adversarial evasion rates while preserving malware functionality &
    Limited to specific Windows PE samples; broader malware families not tested &
    SHAP \\
    \hline
    Ghadekar et al. 2024~\cite{ghadekar2024multi} & 
    Propose a deeperGCN model for multi-class malware detection with integrated GradCAM explainability &
    Reaches 97\% accuracy on combined byte/ASM images; GradCAM highlights key regions &
    Preprocessing (byte/ASM to image) is time-consuming and may limit real-time use &
    GradCAM \\
    \hline
    \end{tabular}}\\
    \vspace{1mm}
    \vspace{-4mm}
\end{table*}

\subsubsection{Gradient-Based Approach}

The gradient-based methodology measures the impact of input features on predictions by assigning weights to different parts of an executable. To elucidate the decision-making processes of Deep Neural Networks (DNNs), Bose et al. \cite{bose2020explaining} examine the MalConv architecture using the open-source 'emberMalConv' framework. This study seeks to understand how the architecture distinguishes between malicious and benign executables based on their raw data. Ember, which is a tool utilized for training static PE malware models within the ML-based domain, highlights the MalConv architecture's ability to attribute significant weight to specific executable parts, thereby having a significant influence on the classification results.

Their research introduces a sophisticated framework based on gradient analysis, which maps gradient embeddings from malicious files and interpolates between accurately classified instances to define a clear decision boundary between categories. By analyzing the interpolation among samples, the study explores filter activations to investigate if there is a connection between different filter pairs. This leads to the development of a correlation heatmap for the filters, providing insights into how they interact. One filter specializes in identifying malicious traits within a file, and another filter focuses on generalizing these findings across various samples. The proposed framework transcends the MalConv model and offers a general method suitable for classification tasks in any neural network.

In summary, gradient-based methods for Windows PE malware detection effectively pinpoint which bytes or segments of an executable are most influential for classification, offering highly granular explanations. However, they often require large labeled datasets and can be susceptible to adversarial manipulation if attackers target the most salient bytes. Despite these limitations, gradient-based XAI remains a powerful tool when fine-grained feature importance is crucial.

\subsubsection{Model-Agnostic-Based Approach}

This method clarifies the predictions by simplifying the complex original model into a more understandable local surrogate model. The research presented by Mathews \cite{mathews2019explainable} introduces an explainability framework aimed at classifying two distinct malware families on Windows PCs. Firstly, they calculate content-based features and extract statistical features derived from Hex and assembly views. These features are indicative of the PE file's structure. Their investigation shows shortcomings in the feature selection process and emphasizes the global characteristics through which a model learns to distinguish between the two malware categories. To elucidate the outcomes produced by the deep learning model, they utilize the LIME framework. 

A study by Pirtch et al.  \cite{pirch2021tagvet} seeks to develop a CNN model that accurately predicts malware tags. This work involves a thorough dynamic analysis that examines malware tags to inform the training of the surrogate learning model. Each detected feature, referred to as a 'token' in the research, receives a relevance score that indicates its impact on the predicted malware tag. To assess the quality of these explanations, the authors employ two measures: descriptive accuracy, which evaluates the precision with which an explanation captures the influential features of a prediction, and descriptive sparsity, which identifies the superfluous features within these explanations. The model's effectiveness is validated through its performance in classifying three types of tags—sandbox, family, and clustering—with each category achieving an accuracy rate of over 90\%. 

In a study, Alani et al. \cite{alani2023xmal} present an ML-based system designed to detect obfuscated malware on Windows platforms with high accuracy and efficiency. The system utilizes a variety of classifiers, including RF, logistic regression, decision trees, Gaussian Naive Bayes (GNB), and extreme gradient boosting (XGB). Through an evaluation process, XGB was identified as the best-performing classifier. Moreover, The system relies on features extracted from memory dumps using the VolMemLyzer tool. The feature selection algorithm, i.e., Recursive Feature Elimination (RFE), identifies the five most effective features, resulting in a streamlined model that maintains an accuracy rate exceeding 99\%. The selected features include the total number of services, average number of dynamic-link libraries (DLLs) per process, total number of mutant handles, number of kernel drivers, and shared process services. The system's detection capabilities are bolstered by its explainability, achieved through SHAP. SHAP values provide insight into the impact of each feature on the model's predictions. The evaluation of the system demonstrates its high accuracy and rapid detection speed, with a processing time of 0.413 microseconds per instance. Despite its robustness, the paper has some potential limitations, such as the model's dependence on specific features and vulnerability to adversarial attacks.

Anthony et al. \cite{anthony2024explainable} focus on enhancing malware detection through the integration of XAI. The primary goal is to address the limitations of traditional ML models, particularly their lack of interpretability. The proposed solution leverages LENs, which offer a balance between accuracy and explainability. LENs provide explanations in the form of First-Order Logic (FOL) rules, making their decision-making processes more transparent and understandable for human analysts. The methodology involves extending the application of LENs to the EMBER dataset. Additionally, they introduce a tailored version of LENs to enhance the fidelity of logic explanations. The experimental results demonstrated that LENs achieve robust performance, rivaling traditional black-box models while significantly outperforming other interpretable methods. The tailored LENs provide high-fidelity explanations with low complexity that can ensure they are both accurate and comprehensible. 

Gulmez et al. \cite{gulmez2024xran} present an approach to ransomware detection by integrating multiple dynamic analysis features with DL and XAI techniques. They developed XRan, which is a system that combines API call sequences, DLL sequences, and mutual exclusion (Mutex) sequences to provide a comprehensive view of executable behaviors. These features are extracted through dynamic analysis, where executables are run in a controlled environment to observe their actions. XRan leverages a two-layer CNN to process these combined sequences, which enables precise detection of ransomware. To address the challenge of model interpretability, the authors integrated two XAI models, i.e., LIME and SHAP.

The study utilized five datasets: RD1 from VirusShare with 6,263 ransomware samples, RD2 from Sorel-20M with 7,703 ransomware samples, RD3 from ISOT with 668 ransomware samples, MD from VX Heaven with 6,263 malware samples, and BD from various sources including Windows System Files and Download.com with 14,797 benign samples. Dynamic analysis was conducted using Cuckoo Sandbox to extract features, which were then combined into sequences for the CNN model. Performance metrics included accuracy, TPR, FPR, and F-score, with XRan showing superior results compared to baseline and state-of-the-art methods.  The experimental results demonstrate XRan's effectiveness, achieving up to a 99.4\% True Positive Rate (TPR) and outperforming existing state-of-the-art methods.

Aryal et al. \cite{aryal2024explainability} aims to enhance the effectiveness of adversarial evasion attacks on malware detectors. They focus on Windows PE malware and utilize SHAP values to identify the most critical regions of malware files that influence detection decisions by a CNN-based malware detector, MalConv. The rationale behind this approach is that by understanding which parts of the malware file have the greatest impact on the detector’s decision, they can strategically place perturbations in these regions to evade detection more effectively.

To achieve this, they calculate the SHAP values for each byte in the malware files using the DeepExplainer module, which is adapted to work with the embedding layer in MalConv. These SHAP values reveal the contribution of each byte to the malware detector's decision, which facilitates the mapping of these values to different regions of the PE file structure. Aggregating these values will help identify the regions with the highest impact. Using this information, they inject adversarial perturbations into these targeted regions, both at a high level (across entire sections) and at a more granular level (within subsections of larger sections). The results, based on a dataset of 6000 Windows PE malware samples, demonstrate that perturbations guided by SHAP values significantly improve the success rate of evasion attacks compared to random perturbations. Specifically, they observe high evasion rates when perturbations are injected in regions with high SHAP values, which demonstrates the efficacy of their explainability-guided approach in crafting adversarial samples that maintain the malware’s functionality while evading detection.

Overall, model-agnostic frameworks (e.g., LIME, SHAP, LENs) are popular for Windows PE malware detection thanks to their flexibility: they can explain virtually any classifier. Yet, these post-hoc explanations can vary from sample to sample, sometimes lacking global consistency. Their simplicity and model independence, however, make them valuable for real-world malware detection pipelines.

\subsubsection{Image-Based Approach}

Recent advancements in CNNs showcase their remarkable capability in detecting malware binaries through image classification techniques. The work presented by Marais et al. \cite{marais2021malware} introduces detection models that effectively convert binary files into grayscale images. Utilizing the Ember dataset, which is formatted in Windows PE, the authors proceed with feature extraction from these grayscale images. Subsequently, they propose a CNN model that leverages these images for malware detection. Additionally, they implement a novel approach, termed the HIT method, to train another CNN model on RGB images. A significant contribution of their research is the application of the GradCam++ explainability technique on the CNN model. This technique identifies the most influential pixels affecting the model's prediction, aiming to diminish the false positive rate of detecting malicious files.

Contrastingly, while non-linear ML models are known for their superior accuracy and classification performance over linear counterparts, their complexity often renders them difficult to interpret. Addressing this challenge, Li et al. \cite{li2021effectiveness} have developed an IFFNN. This model can achieve high accuracy in malware detection and ensures interpretability. They conduct their experiments on a Windows server to determine the IFFNN's capability to handle multi-class classification problems. Moreover, to assess the effectiveness and interpretability of the IFFNN, they use the MNIST dataset for image classification and convolutional layers for a comprehensive qualitative evaluation of the model's interpretability.

In another work, Chen et al. \cite{chen2019believe} aim to enhance the interpretability of image-based dynamic malware classification by extending the LIME framework. They start by training deep learning models on images and then apply an explanatory approach to understanding the decision-making process of these models. The objective is to determine whether the insights derived from the algorithm align with expert knowledge in the cybersecurity domain.

Lin and Chang \cite{lin2021towards1} engage with an image-based malware dataset to explore the potential of ensemble learning. They introduce a Selective Deep Ensemble Learning (SDEL)-based detector that is coupled with an innovative Interpretable Ensemble Learning approach. This detector is specifically designed for Malware Detection (IEMD). The IEMD strategy is developed to elucidate the predictive decisions made by the SDEL detector and advance the interpretability of the model. This endeavor is supported by the deployment of explainable AI techniques such as LIME, SHAP, and Layer-wise Relevance Propagation (LRP). These methods are analyzed and compared to understand their efficacy in providing transparent explanations. Their research results have impressive outcomes, achieving an accuracy rate of approximately 99.87\%. Furthermore, the study shows the superiority of their explanations in the context of image-based malware classification compared to preceding research.

The paper by Ciaramella et al. \cite{ciaramella2024explainable} develops an approach to ransomware detection by converting Windows PE files into RGB images and analyzing them using DL-based models. The researchers developed a script to transform the binary code of executable applications into images, which are then used as input for various CNNs such as LeNet, AlexNet, Standard-CNN, and VGG-16. The goal is to classify the files into ransomware, generic malware, or legitimate software. The Grad-CAM technique is employed to enhance the interpretability of the model’s predictions. Grad-CAM generates visual explanations by highlighting regions of the images that most influence the model's decisions. The results demonstrate the effectiveness of the proposed method, achieving high accuracy, precision, and recall, particularly with the VGG-16 model, which outperformed others with an accuracy of 96.9\%.  

Ghadekar et al. \cite{ghadekar2024multi} implement a methodology for detecting various types of malware by leveraging a modified GNN architecture called deeperGCN, along with XAI techniques. The research combines byte and ASM (assembly) files, converting them into images to better capture intricate malware behaviors. This conversion process involves the extraction of features such as byte bigrams, opcode sequences, and the generation of pixel representations of the files. These images are processed using the deeperGCN model, which enhances the feature extraction capabilities by leveraging the inherent relationships in the graph-structured data. The model includes several advanced techniques, such as skip connections to address vanishing gradient problems and a graph readout pooling layer to effectively aggregate information across nodes.

The results demonstrate that this innovative approach achieves a high detection accuracy of up to 97\%. In addition, the integration of GradCAM provides transparency into the model's decision-making process by generating heatmaps that highlight the important regions of the input data influencing the predictions.

By converting PE files into images, these techniques leverage CNNs’ strength in pattern recognition. Gradient-based visualization methods (e.g., Grad-CAM) then highlight which pixel regions most impact the model’s decision, increasing transparency for security analysts. The main challenge is the computational cost and the complexity of transforming binaries to images, but when accuracy and visual interpretability are desired, image-based XAI can be highly effective.

Collectively, these Windows-based XAI approaches illustrate a variety of explainability techniques—gradient-focused, model-agnostic, and image-based—each balancing interpretability, detection accuracy, and computational complexity. Next, we explore how Android malware detection demands similar yet distinct methods, given Android’s unique environment and features.

\subsection{Android-Based Malware Approaches}

\begin{table*}
    \centering
    \footnotesize
    \setlength{\tabcolsep}{0.8\tabcolsep}
    \def\arraystretch{1.5}
    \caption{Research Addressing Explainable Machine Learning in Android Malware}
    \resizebox{\textwidth}{!}{%
    \begin{tabular}{|p{1.25cm}|p{3.75cm}|p{5.5cm}|p{3.5cm}|p{2cm}|}
    \hline
    \textbf{Paper/Year} & \textbf{Focus/Objective} & \textbf{Contribution} & \textbf{Limitation} & \textbf{XAI Technique}\\
    \hline
    Kumar et al. 2018~\cite{kumar2018effective} & Reduce high-dimensional Android malware features to avoid unnecessary data &
    Proposed static analysis + feature vectorization for effective Android malware detection &
    Relies on small static datasets; lacks hybrid or dynamic analysis &
    Feature Extraction \\
    \hline
    Melis et al. 2018~\cite{melis2018explaining} &
    Highlight global features to differentiate benign from malicious Android apps &
    Presented a gradient-based, global surrogate method to explain black-box Android malware models &
    Does not compare different surrogate models’ impact on explanations &
    Global Surrogate \\
    \hline
    Iadarola et al. 2021~\cite{iadarola2021semi} & 
    Evaluate deep learning for Android malware family identification &
    Introduced Grad-CAM-based CNN to localize crucial image regions for malware family classification &
    Susceptible to code obfuscation &
    Grad-CAM \\
    \hline
    Kinkead et al. 2021~\cite{kinkead2021towards} &
    Compare CNN-identified malicious opcode sequences with LIME explanations &
    Introduced a CNN to locate malicious opcode sequences; validated with LIME consistency &
    No deep analysis of CNN filter activations &
    LIME \\
    \hline
    Lu and Thing 2021~\cite{lu2021does} &
    Address Android malware detection via feature attribution &
    Proposed MPT explainer to optimize feature attribution; compared with LIME/SHAP &
    Limited scope on a single attack; no mitigation strategies &
    Modern Portfolio Theory (MPT) \\
    \hline
    Korine and Hendler 2021~\cite{korine2021daemon} &
    Present DAEMON, a model-agnostic, explainable malware classifier &
    Developed DAEMON with layer-wise propagation for explainable classification across platforms &
    Unclear performance on benign executables &
    Feature Importance \\
    \hline
    Scalas 2021~\cite{scalas2021malware} &
    Explore feature-based traits for effective Android ransomware detection & 
    Proposed a gradient-based approach and analyzed adversarial attacks with integrated gradients &
    Lacks detailed rationale for specific attribution methods &
    Integrated Gradient \\
    \hline
    Yan et al. 2021~\cite{yan2021effective} &
    Introduce a rule-extraction method from DNNs for transparent mobile malware detection &
    Extracts DNN rules for malicious traffic detection; offers high accuracy and explainability &
    Lacks detail on the proposed online detection implementation &
    Rule-Based Learner \\
    \hline
    Wang et al. 2016~\cite{wang2016trafficav} &
    Leverage network traffic for mobile malware detection; add interpretable ML explanations &
    Developed TrafficAV with feature extraction + ML classification for malware traffic &
    Small dataset and limited model comparisons &
    Feature Extraction \\
    \hline
    Iadarola et al. 2021~\cite{iadarola2021towards} &
    Propose explainable DL for mobile malware detection and family classification &
    Implemented Grad-CAM to highlight crucial image areas for accurate classification &
    Potential label bias in the dataset &
    Grad-CAM \\
    \hline
    Wu et al. 2021~\cite{wu2021android} &
    Combine MPT explainer (XMAL) with feature descriptions for high-fidelity Android malware classification &
    Employs XMAL for interpretable detection, validated via user surveys and state-of-the-art comparison &
    Missing multi-attention; limited feature scope &
    XMAL, LIME \\
    \hline
    Alenezi and Ludwig 2021~\cite{alenezi2021explainability} & 
    Apply SHAP for explainable cybersecurity threat detection &
    Tested RF, XGBoost, and sequential models; SHAP reveals key features &
    Limited model variety tested &
    SHAP \\
    \hline
    Ullah et al. 2022~\cite{ullah2022explainable} & 
    Develop BERT-based transfer learning for Android malware, with SHAP for top features &
    Merged text (BERT) and image-based (CNN) features; explained predictions with SHAP &
    Only SHAP used, no LIME comparisons; local explanations only &
    SHAP \\
    \hline
    Naeem et al. 2022~\cite{naeem2022explainable} & 
    Fine-tune CNN for IoT malware images, offering Grad-CAM-based explainability &
    Applied Inception-v3 for IoT malware, employing Grad-CAM heatmaps for explanation &
    Limited validation and scalability to larger malware families &
    Grad-CAM \\
    \hline
    Alani and Awad. 2022~\cite{alani2022paired} & 
    Introduce PAIRED, a lightweight Android malware detector with SHAP feature importance &
    Reduced features from 214 to 35, achieving ~98\% accuracy; SHAP explains global feature impact &
    Only global SHAP used, no local interpretation &
    SHAP \\
    \hline
    Liu et al. 2022~\cite{liu2022explainable} & 
    Examine temporal bias in Android malware detection and use XAI to explain inflated accuracy &
    Reveal how inconsistent time splits inflate ML performance, with XMal and Drebin for explanation &
    Feature-importance XAI may miss deeper non-linear interactions &
    XMAL, DREBIN, and Model-Agnostic Techniques \\
    \hline
    Ambekar et al. 2024~\cite{ambekar2024tablstmnet} & 
    Propose TabLSTMNet combining TabNet + LSTM for interpretable Android malware detection &
    Achieves ~97–98\% accuracy, with LIME/SHAP explaining permission-level contributions &
    Over-sampling risks overfitting; high computational complexity &
    LIME, SHAP \\
    \hline
    Soi et al. 2024~\cite{soi2024enhancing} & 
    Use function-call graphs and SHAP for clear, model-agnostic explanations in Android &
    Selects critical API calls from function-call graphs, employs SHAP for transparent classification &
    Large-scale API features may hinder efficient analysis &
    SHAP \\
    \hline
    \end{tabular}}\\
    \vspace{1mm}
    \vspace{-4mm}
    \label{tab:Android}
\end{table*}

The rapid advancement of technology has also led to an increase in malware attacks, with the Android platform emerging as a particularly significant target. In response to this escalating threat, various security measures have been implemented within the Android ecosystem. Among these, ML-based methods have proven to be highly effective in detecting Android malware, which led to extensive research in this domain \cite{sahs2012machine} \cite{xiao2019android} \cite{demontis2017yes} \cite{yan2018lstm}. Specifically, recent developments in DNNs have improved detection rates and reduced the reliance on manual feature engineering.

A standout innovation in this field is the DREBIN malware detection system \cite{arp2014drebin}. DREBIN leverages a lightweight approach to identify Android malware on smartphones through static analysis, extracting application features that are represented in a binary vector format. This setup enables linear classification to differentiate between features of benign and malicious applications. Furthermore, DREBIN is distinguished by its explainable approach to malware detection. It provides insights into the reasoning behind its decisions by highlighting key attributes of detected malware.

The dataset used by DREBIN includes 5,560 malware samples and 123,453 benign samples that demonstrate the comprehensive nature of its analysis.  DREBIN has outperformed other ML-based approaches by achieving high accuracy. This success has attracted significant attention in the academic world and has prompted many researchers to leverage DREBIN in their studies on explainable Android malware detection. This has made a substantial contribution to the enhancements of mobile security.

To analyze the Android malware, the research by Kumar et al. \cite{kumar2018effective} introduces two ML-supported methodologies: one focuses on static analysis and the other on feature extraction. They perform feature extraction on the DREBIN dataset through vectorization, followed by feature selection and dimensionality reduction, thus transforming high-dimensional data into a more manageable low-dimensional format while omitting extraneous features. Their analysis yields metrics such as the True Positive Rate (TPR) and False Positive Rate (FPR), with their methodology demonstrating high precision and recall. Various ML-based algorithms, including SVM, KNN, Naive Bayes, and C4.5, are applied to the newly processed data, which reveals  SVM's superior performance. The combination of static analysis, feature vectorization, and supervised learning enables these ML algorithms to identify new malware families with high true positive and recall rates.

While feature extraction and dimensionality reduction can streamline malware detection, filtering out redundant instances in the dataset can further enhance both classifier efficiency and the clarity of explanations. Surendran et al. \cite{surendran2024optimizing} propose using the Ochiai coefficient to identify and remove near-duplicate samples before retraining, which can reduce noise and training overhead. Eliminating these overlaps helps ensure that model explanations (e.g., SHAP or LIME outputs) capture truly informative patterns rather than repeated artifacts, ultimately improving both performance and the stability of XAI-based insights.

Following this discussion, below we will explore gradient-based approaches, which provide a more fine-grained analysis of feature importance in malware classification.

\subsubsection{Gradient-Based Approach}

Gradient-based methods are crafted to classify and detect malware through the lens of ML. These strategies reveal the underlying architecture of a specific ML model and enhance the interpretability of predictions made by deep learning-based malware detection systems. This approach calculates and allocates the predictive weights relative to input features across different segments of the executable file. To discover the mechanism behind black-box Android malware detection systems and determine the most significant features influencing each decision, Melis et al. \cite{melis2018explaining} introduce a comprehensive explainable ML framework. This framework uses a gradient-based technique to determine whether a sample is correctly classified as malware, leveraging its most critical local features. Their research utilizes the DREBIN Android malware detection tool for practical testing. The main objective of this endeavor is to enhance the accuracy of predictions while maintaining the transparency and interpretability of the decision-making process. By utilizing the DREBIN malware detection tool and dataset, the authors propose a novel methodology that highlights both local and global characteristics to distinguish and clarify the discernment between benign and malicious applications.

The authors, Iadarola et al. \cite{iadarola2021semi}, introduce a gradient-based deep learning methodology designed to clarify the methodology behind malware family classification. This approach begins by extracting code from Android application package (.apk) samples and subsequently transforming it into an image format. Following this transformation, a CNN model classifies these images into their respective malware families. They implement the Gradient-weighted Class Activation Mapping (Grad-CAM) technique to facilitate the prediction of classes by identifying critical areas within the images. A thorough code analysis is conducted to demonstrate the efficacy of their method in extracting relevant classes. In another work, Scalas \cite{scalas2021malware} develops a gradient-based strategy specifically designed for detecting Android ransomware. This study shows the selection of system API calls as key features, asserting their utility in evading detection strategies employed by attackers.

Further study by Melis et al. \cite{melis2022gradient} explores the effectiveness of gradient-based attribution techniques in identifying key features crucial for understanding a classifier's decision-making process. Their work seeks to establish the importance of these features in developing more robust algorithms. They analyze the correlation between explanatory methods and adversarial resilience, probing how these aspects are interconnected. Moreover, Iadarola et al. \cite{iadarola2021towards} put forward an explainable deep-learning framework designed for mobile malware detection. This approach transforms applications into images that feed into an explainable deep-learning model that is capable of recognizing Android malware and classifying its family. Utilizing the Grad-CAM explainability method, they demonstrate the selection of explanatory techniques that improve classification performance. To enhance interpretability, they generate heatmaps that offer visual insights into the model's reasoning, making the predictions' rationale more accessible. Additionally, because the process of analyzing these heatmaps is automated, it simplifies the architecture's debugging for analysts without necessitating a background in the system's design. besides enhancing transparency in their model, they record a notable increase in accuracy.

Naeem et al. \cite{naeem2022explainable} expand the application of gradient-based methods by introducing a transfer learning approach for classifying IoT malware, leveraging the Inception-v3 architecture, a pre-trained network designed to process malware images and extract pivotal features. These features are then fed into a classification algorithm and evaluated across various ML classifiers to assess performance. The Grad-CAM explainability method is utilized to highlight the critical areas within the images. Furthermore, the study utilizes t-distributed stochastic neighbor embedding (t-SNE) to verify the comprehensiveness of the feature set within the proposed CNN models. This ensures that they encapsulate sufficient information for effective malware classification.

Gradient-based attribution helps Android malware detectors reveal exactly which app features (e.g., API calls, opcode sequences) most strongly affect the classifier’s output. While this granularity aids in pinpointing malicious behavior, these methods may require careful tuning to handle the massive variety of Android apps and can still be undermined by obfuscation or adversarial feature manipulation.

\subsubsection{Model-Agnostic Based Approach}

The challenge of explaining the vast range of models in deep learning research is increasing. In this context, model-agnostic approaches provide explanations after the decision-making process, which are applicable to various opaque models. The study by Kinkead et al. \cite{kinkead2021towards} presents a novel CNN-based method focused on identifying specific parts of opcode sequences suspected of containing malicious elements. Their main objective is to examine and compare the similarities between the locations of malicious opcode sequences identified by the CNN and those marked as important by LIME. They carry out their research using the DREBIN dataset, known for its collection of 5,560 malicious apps across different malware families, serving as a standard for Android malware detection. Their results show that the model achieves an accuracy of about 0.98, highlighting CNN's exceptional performance with the DREBIN dataset. Further analysis of how both CNN and LIME highlight locations across all the samples in each malware family reveals a significant finding that CNN tends to focus on the same areas as LIME, indicating CNN's targeted effort in detecting malware.

In another effort to improve Android malware detection, researchers Lu and Thing \cite{lu2021does} utilize a model-agnostic explainable AI framework focused on feature attribution, highlighting the importance of feature manipulation and optimization. Their approach integrates a trained model with a Modern Portfolio Theory (MPT) explainer during the explanation phase. Quantitative analysis of their method shows greater sensitivity in detecting important data features compared to the results from machine learning-based Android malware detection tools. Additionally, they use both LIME and SHAP to evaluate the effectiveness of the MPT explainer, seeking to confirm its superior capability in identifying key features essential for malware detection.

On the other hand, rapidly mutating malware variants necessitate sophisticated classification methods to categorize these variants accurately. Although variants within the same malware family often exhibit identical behavioral patterns, the increasing number of variants complicates the process of accurately classifying new ones. This challenge has motivated researchers to develop advanced detection tools aimed at enhancing the accuracy of malware classification. One notable contribution in this field is DAEMON, a data-agnostic malware classification tool developed by Korine and Hendler \cite{korine2021daemon}. DAEMON stands out for its ability to discern the unique features of various malware families, which lends clarity and explainability to the classification process. The researchers behind DAEMON have collected extensive datasets, which they have analyzed on both Windows and Android platforms, utilizing the renowned DREBIN dataset for the latter. Their efforts have culminated in DAEMON achieving remarkable accuracy in malware classification.

Based on a model-agnostic approach, the study \cite{ullah2022explainable} introduces a novel, hybrid methodology for crafting an explainable malware detection system that leverages both textual and visual representations of malware attributes. Initially, they develop a pre-trained model known as Bidirectional Encoder Representations from Transformers (BERT), specifically customized to learn textual features derived from network traffic. Following this, they suggest an algorithm capable of transforming malware into a visual format. Subsequently, a CNN model is implemented to utilize deep extraction of features. Once these balanced features are obtained, they are fed into a suite of ensemble models, including SVM, DT, LR, and RF, to facilitate the system's classification and detection capabilities. Furthermore, the researchers utilize SHAP, a model-agnostic technique for explainability, to elucidate the critical features in interpreting the model’s decisions.

In the study Alani et al. \cite{alani2022paired}, the researchers introduce an Android malware detection system named PAIRED. This system is distinguished by its lightweight design and high precision, achieving a significant reduction in feature count—from 214 to 35, amounting to an approximate 84\% decrease. The SHAP explainability technique is utilized to elucidate the overarching influence of the features, identifying which among them have a greater impact on the predicted outcomes. Impressively, PAIRED manages to sustain a remarkable accuracy rate of 97.98\%.

Ambekar et al. \cite{ambekar2024tablstmnet} introduce the TabLSTMNet, an approach to Android malware classification that combines the strengths of the TabNet architecture, which was developed by researchers at Google Cloud and LSTM models, complemented by XAI techniques. This model integrates TabNet’s attention-mechanism feature selection, which efficiently identifies critical features, with LSTM's dynamic processing capabilities for sequential data. This integration allows for a detailed analysis of Android permissions and API calls to distinguish between benign and malicious applications effectively. The proposed model is evaluated on two different datasets and achieves classification accuracies, demonstrating 97.10\% on the NATICUSdroid dataset and 98.00\% on the TUNADROMD dataset. Moreover, the incorporation of explainable AI methods such as LIME and SHAP significantly increases the transparency of the model’s decision-making process.

Soi et al. \cite{soi2024enhancing} propose a novel methodology for improving the explainability of Android malware detection systems. The approach begins with a static analysis of Android application packages (APKs) to extract a Function Call Graph (FCG). This graph represents all the API calls within the application's code. Based on FCG, a set of critical API calls can be selected, which are strongly correlated with the application's behavior, to serve as features for their model. After that, the selected features are embedded using Natural Language Processing (NLP) techniques, such as TF-IDF and Word2Vec, to produce a consistent input format for a CNN. To enhance the interpretability of the model's decisions, the paper employs SHAP values, which provide a clear and detailed explanation of how each API call contributes to the classification outcome.

The results of the experiments conducted on a dataset of over 40,000 Android applications show that the proposed method achieves a classification accuracy comparable to state-of-the-art models. The paper also conducts extensive evaluations to address potential issues such as temporal bias and concept drift. It highlights that while the approach maintains strong performance over time, the inclusion of more recent data can be important for sustaining its accuracy. 

Model-agnostic tools like LIME and SHAP allow security analysts to investigate any black-box Android malware detector, highlighting the top features—like permissions or API calls—responsible for a malicious label. Their interpretability fosters user trust but can sometimes yield inconsistent local explanations across app variants, especially if the underlying model is unstable.

\begin{table*}
    \centering
    \footnotesize
    \setlength{\tabcolsep}{0.8\tabcolsep}
    \def\arraystretch{1.5}
    \caption{Research Addressing Explainable Machine Learning in Malware Analysis.}
    \resizebox{\textwidth}{!}{%
    \begin{tabular}{|p{1.25cm}|p{3.5cm}|p{5.5cm}|p{3.5cm}|p{2.25cm}|}
    \hline
    \textbf{Paper/Year} & \textbf{Focus/Objective} & \textbf{Contribution} & \textbf{Limitation} & \textbf{XAI technique}\\
    \hline
    \multicolumn{5}{|l|}{\textbf{Hardware-Based Malware Approaches}}\\
    \hline 
    Pan et al. 2020~\cite{pan2020hardware} &
    Use linear regression for interpretable hardware-based malware detection &
    Introduced a hardware trace–based detection framework with interpretable regression &
    Insufficient details on filter/activation interpretability &
    Linear regression \\
    \hline
    Pan et al. 2022~\cite{pan2022hardware} &
    Hardware performance counters + ETB for explainable malware detection/localization &
    Provided interpretable HPC-based classification, localizing malicious behavior via RNN + DT &
    Does not test various surrogates or compare interpretability outcomes &
    Linear regression and Decision tree \\
    \hline
    Li et al. 2021~\cite{li2021mad} &
    Introduce I-MAD with an interpretable feed-forward neural network for hardware-level detection &
    Analyzes assembly code, uses IFFNN to locate payload patterns and improve transparency &
    Limited fidelity documentation; risk of adversarial misuse &
    IFFNN \\
    \hline
    \multicolumn{5}{|l|}{\textbf{Linux-Based Malware Approaches}}\\
    \hline 
    Wang et al. 2021~\cite{wang2021effective} & 
    Pinpoint malicious code snippets in Linux malware with LRP-based explainability &
    Proposed a CNN to detect inline assembly malware, using LRP to highlight influential code blocks &
    Minimal discussion of future improvements &
    Layer-wise relevance propagation (LRP) \\
    \hline
    Wang et al. 2021~\cite{wang2021exposing} & 
    Develop SHAP-guided adversarial evasion on Linux malware detectors &
    Combined feature-level and problem-space obfuscation using SHAP-based insights &
    Limited to static detection and SHAP approach &
    SHAP \\
    \hline
    Mills et al. 2019~\cite{mills2019efficient} & 
    Propose NODENS, an interpretable RF-based real-time Linux malware detector &
    Demonstrates zero-day detection with a tree-based approach for straightforward explanations &
    Evaluation constrained by small dataset &
    Decision tree \\
    \hline
    \multicolumn{5}{|l|}{\textbf{PDF-Based Malware Approaches}}\\
    \hline
    Kuppa and Le-Khac 2020~\cite{kuppa2020black} & 
    Explore black-box attacks on gradient-based XAI, using PDF and Android malware &
    Investigated gradient-based XAI robustness and accuracy under black-box adversarial scenarios &
    No detailed defense strategies were provided &
    Gradient-Based Methods \\
    \hline
    Severi et al. 2021~\cite{severi2021explanation} & 
    Demonstrate SHAP-based backdoor triggers that poison ML training &
    Created stealthy backdoor triggers for PDF/Windows PE; validated with SHAP analysis &
    Constrained to certain backdoor scenarios &
    SHAP \\
    \hline
    Guo et al. 2018~\cite{guo2018lemna} & 
    Introduce LEMNA for interpretable PDF malware classification with fused lasso &
    LEMNA captures feature dependencies, outperforming LIME on PDF malware tasks &
    Limited detail on feature-level explanation &
    LEMNA \\
    \hline
    \end{tabular}}\\
    \vspace{1mm}
    \vspace{-4mm}
    \label{tab:Hardware/PDF/Linux}
\end{table*}

\subsubsection{Rule-Based Approach}

Yan et al. \cite{yan2021effective} present an innovative approach for extracting rules from DNNs, aiming to balance the accuracy intrinsic to DNNs with the need for explainability in their operation.  The initial phase contains the collection of network traffic data, utilizing a tool named DroidCollector for this purpose. Subsequent to data collection, feature extraction is conducted to distill the essential information necessary for training the model. This algorithm begins by verifying the appropriateness of the neural network settings, such as its suitability for classification tasks. In instances where the predicted label aligns with the true label, the model's performance is deemed satisfactory. However, misclassification triggers a reassessment and update of the neural network's weights. The backpropagation process prioritizes the weights of the outermost layer before sequentially addressing each subsequent layer, effectively distributing the errors from each output variable across the network's hidden layers. Through numerous iterations, the model iteratively refines itself until it achieves optimal performance, at which point rules are extracted from the DNN.

Employing these extracted rules, the authors devise a mechanism to detect malicious network activity. They conduct an evaluation of their DNN rule extraction technique against three contemporary technologies—MultiView, CNN, TrafficAV—and four ML-based algorithms, namely Bagging, Adaboost, KNN, and Random Forest. The findings from this comparison suggested the superiority of their proposed method, which excelled in numerical prediction accuracy and outperformed the benchmarked methods. Hence, the authors propose an online detection system that is optimized for high-speed network environments and leverages FPGA technology to facilitate the real-time detection of mobile malware. 

Also, as ML has emerged as a powerful tool for uncovering rules crucial for predictive data analysis, in their work, Wang et al. \cite{wang2016trafficav} develop TrafficAV, an efficient and intelligible method for classifying mobile malware based on network traffic patterns. This approach is designed to minimize resource usage and performs malware detection and network traffic analysis server-side. TrafficAV leverages feature extraction combined with the C4.5 DT algorithm to detect the presence of malicious applications, applying two distinct detection models for HTTP and TCP protocols. This dual-model strategy has yielded high accuracy rates. Furthermore, TrafficAV provides an analysis of the significance of each feature in the decision-making process, offering user-friendly explanations of its findings.

Similarly, another study  \cite{wu2021android} works on the framework named XMAL—an interpretable machine learning-based framework—proposes a rule-based methodology for accurately classifying Android malware. This system enhances its capabilities with a Multilayer Perceptron (MLP) model that incorporates an attention layer to highlight the relevance of input features. By integrating the MLP model, the researchers are able to underscore the importance of specific features in malware identification. XMAL's effectiveness is also compared to other explainability techniques, such as LIME, where it demonstrates superior performance in terms of interpretability, thereby reinforcing the value of machine learning in enhancing cybersecurity measures.

The study provided by Liu et al. \cite{liu2022explainable} investigates the performance of ML models under realistic and unrealistic experimental setups. It utilizes a dataset of 165,000 Android applications, with 33,000 malware and 132,000 benign samples, spanning from 2010 to 2020. The focus of the study is on understanding why ML-based malware detection models perform exceptionally well under certain experimental setups, particularly those involving temporal inconsistencies between malicious and benign samples. To achieve this goal, it leverages different explainable malware detection techniques, i.e., XMal, Drebin, and model-agnostic explanation approaches \cite{fan2020can}. It shows how this inconsistent distribution between malware and benign samples can lead to high detection performance but poor generalization. Its results emphasize the need for XAI techniques in experiments to ensure that the models are practically useful and not just theoretically effective.

Rule-based learners (e.g., DNN rule extraction or tree-based methods) strive to combine the accuracy of deep nets with the readability of logical rules. When applied to Android malware, these frameworks can yield straightforward if-then statements describing malicious behaviors. Still, as Android evolves, rule sets must be updated frequently, or attackers may exploit static rules for evasion.

Collectively, these Android XAI approaches underscore the interplay between interpretability and the fluid nature of the mobile ecosystem, where APIs and permissions often shift. Next, we examine how hardware-based XAI approaches address embedded systems and HPC-driven malware detection.

\subsection{Hardware-Based Malware Approaches}

Many researchers have designed hardware-based malware detectors with the assumption that solutions such as anti-virus software can be fooled easily by malicious code.  Although there are many studies on hardware-assisted malware detection, there is a distinct lack of research focused on explainability. This highlights a critical need for innovation in the field that goes beyond detection accuracy, aiming for systems that can articulate the rationale behind their detections. In table \ref{tab:Hardware/PDF/Linux}, the literature on hardware malware classification and detection using XAI is discussed.  

Sheldon \cite{pan2021hardware} conducts a study analyzing hardware traces for malware detection using explainable ML approaches. Hardware traces comprise the data stored in caches and registers during the execution of a program. These traces are then processed by an ML-based model to identify malware presence. Subsequently, the model's accuracy is evaluated against that of leading-edge ML models. The feedback obtained from this comparison is utilized to refine the accuracy of malware detection.

\begin{figure}
    \centering
    \includegraphics[width=\columnwidth]{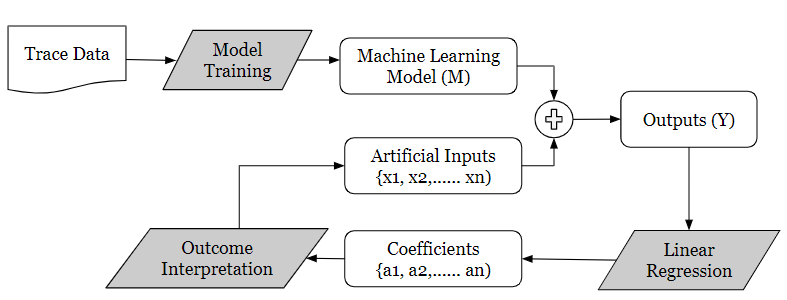}
    \caption{Explainable Hardware Malware generation workflow \cite{pan2020hardware}.}
    \label{fig:Hardware Malware Approach}
\end{figure}

To explore the comprehensive study of hardware trace analysis and the development of an explainable, hardware-assisted malware detection framework, the research by Pan et al. \cite{pan2020hardware} introduces a hardware framework depicted in Fig \ref{fig:Hardware Malware Approach}. This framework is structured around three principal activities.

The first phase involves training an ML model (M) with collected hardware traces. For this purpose, they implement an RNN and leverage the Embedded Trace Buffer (ETB) architecture for trace collection. Subsequently, a specially curated artificial dataset $X = (x_1, x_2, \dots, x_n)$ is processed by the machine learning model (M) to yield the output $Y$. To adapt this artificial dataset for the model, linear regression is conducted, resulting in the formulation of a linear predictive model. This process is shown in Equation \ref{Hardware_equation1}. 

\begin{equation}\label{Hardware_equation1}
    y = \sum_{i=1}^{n} a_i x_i + \epsilon
\end{equation}

Linear predictions are formulated as a polynomial function, where $n$ represents the number of instances. This expression incorporates an error term, $\epsilon$, which is crucial for understanding the weight distribution within the model. The value of $\epsilon$ needs to be as small as possible. The goal is to minimize the value of $\epsilon$ as much as possible, ensuring that the model's predictions are as accurate and reliable as possible.

\begin{equation}\label{Hardware_equation2}
    \arg \min ||X_a - y||_2
\end{equation}

Equation \ref{Hardware_equation2} shows the optimization problem, emerging from the selection of $y$ as the perturbed output. Further ridge regression is applied in order to achieve higher fitness with correlated data. This approach aims at achieving optimal fitness by introducing an additional term to the optimization equation, further elaborated in Equation \ref{Hardware_equation3}.    

\begin{equation}\label{Hardware_equation3}
    \arg \min ||X_a - y||_2 + \lambda||a||_2
\end{equation}

To mitigate the issue of high variance, the strategy involves substituting $X$ with $X - \lambda I$, as depicted in Equation \ref{Hardware_equation4}. This adjustment incorporates a regularization parameter, $\lambda$, and the identity matrix, $I$, directly into the predictive model. This technique effectively reduces the model's complexity, discouraging overfitting by penalizing larger coefficients. 

\begin{equation}\label{Hardware_equation4}
    \arg \min ||X_a - y||_2 + \lambda||a||_2 \rightarrow \arg \min ||(X - \lambda I)a - y||_2
\end{equation}

After determining the linear regression coefficients, the focus shifts to interpreting the outcomes, with particular emphasis on identifying the most influential features. Features associated with larger coefficients are flagged as potentially malicious. To benchmark the effectiveness of their method, the authors reference the PREEMPT malware detector for comparison \cite{basu2019preempt}. PREEMPT employs two algorithmic models, i.e., Random Forest and Decision Tree, utilizing hardware performance counters (HPC) to generate its dataset. However, it does not extensively investigate the dataset's characteristics, focusing primarily on the interpretation of outcomes. Given the scarcity of research in this area and the need to address malware detection challenges, it is imperative for researchers to engage deeply with the realm of explainable hardware malware detection.

Building upon their previous efforts, the authors \cite{pan2022hardware} harness hardware performance counters and embedded trace buffers to identify the exact locations of malicious activities within a system. They develop Decision Tree and RNNs to perform a trade-off between accuracy and efficiency in their detection methodology. Their evaluation, conducted on a broad spectrum of real-world malware datasets, elucidates the interpretability of the RNN model, leveraging linear regression and Decision Tree through tree parsing techniques.

In a parallel line of research, Li et al. \cite{li2021mad} introduce an innovative interpretable malware detector named I-MAD transformer, designed to analyze assembly code at the basic block level of executables. This approach integrates an interpretable feed-forward neural network, allowing for the examination of each feature's impact on the prediction outcome. The significant advancements brought forth by this study include 1) The proposition of a deep learning model capable of interpreting entire sequences of assembly-level code in malware executables, offering a comprehensive analysis beyond superficial layers. 2) The introduction of two pre-training activities aimed at enhancing the understanding of the relevance and functionality of assembly-level constructs, thereby improving the model's predictive accuracy and interpretability. 3) The development of an Interpretable Feed-Forward Neural Network (IFFNN), which assists analysts in identifying payloads and recurring patterns within malware samples. This network combines the interpretability akin to logistic regression with the modeling prowess of multi-layer FFNN, presenting a powerful tool for cybersecurity professionals in the battle against malware.

In conclusion, hardware-based XAI solutions exploit HPCs and trace buffers to offer fine-grained inspection of runtime behavior. The interpretable models (e.g., linear regression, interpretable feed-forward nets) provide valuable insights into which low-level counters or basic blocks signal malicious patterns. However, custom hardware or specialized instrumentation is often needed, which can limit widespread deployment.

\subsection{PDF-Based Malware Approach}

A Portable Document Format (PDF) file contains text, images, digital signatures, and other elements. Its structure includes a Header, Body, Cross-reference Table, and Trailer. The header of a PDF file is the top section that indicates the version number and file format. The body of the PDF stores all the pertinent data, and it contains a range of objects, including data, text, images, and dictionaries. The cross-reference table in a PDF file includes links to all elements within the document, which facilitates navigation and access. The trailer, which links to the cross-reference table, also contains the EOF marker.

Given the global acceptance of PDF as a standard document format, the prevalence of PDF malware is increasing. Malware exploits vulnerabilities in PDF readers to hijack execution control, such as executing shell code. To evade malware detection, PDF authors might employ techniques that can cause the PDF reader to crash  \cite{chen2020training}.

To analyze and evaluate the consistency and correctness of the gradient-based approach, the study by Kuppa and Le-Khac \cite{kuppa2020black} designed a novel black-box attack. They apply this method to detect malicious PDF files using the Mimicus dataset and to identify Android malware with the DREBIN dataset, interpreting the results through gradient-based explainable machine-learning techniques.

To guide the selection of relevant features and avoid backdoor poisoning attacks, the authors \cite{severi2021explanation} use model-agnostic techniques in explainable machine learning and develop effective backdoor triggers. They specifically use Android, PDF, and Windows PE files for malware classification and analyze nearly 10,000 samples of benign and malicious files. To maintain the functionality of the binaries, they create a static analysis watermarking utility for Windows PE files that meets multiple adversarial constraints. Subsequently, their attention turns to PDF files and Android applications. Using the SHAP explainability technique, they identify features that contribute to malware detection. Finally, they demonstrate and evaluate the challenges in fully defending against these stealthy poisoning attacks.

To classify PDF malware, another research \cite{guo2018lemna} introduces an explainable method named LEMNA, which provides high-fidelity explanations for malware detection. They utilize deep learning models and assess their interpretability using LEMNA. The study also explores feature augmentation, along with synthetic and feature deduction tests. They note that due to sparse input feature vectors affecting local decision boundaries, LIME, and other advanced explainable techniques were as ineffective as traditional feature selection methods. The PDF-based malware detection studies are summarized in Table \ref{tab:Hardware/PDF/Linux}.

PDF files continue to be a popular vector for malicious exploitation. The interpretability methods described here—whether gradient-based or model-agnostic—provide clarity on which structural elements (e.g., JavaScript objects, suspicious headers) are raising red flags. Nevertheless, sophisticated obfuscation within PDFs can still pose significant challenges for these models.

\subsection{Linux-Based Malware Approach}

Linux, a Unix-based operating system, is an open-source platform that is renowned for its reliability and functionality. For malware analysis, Linux allows malicious code to run in isolated sandbox environments. However, due to the limited availability of sandboxes compatible with the latest Linux versions, they are less commonly used than Android and Windows platforms. Recognizing that ML in malware detection often yields predictions that lack explainability, Wang et al. \cite{wang2021effective} introduce an explainable malware detection method based on Linux systems. This approach clarifies the rationale behind the classifier's decisions by locating the malicious code snippets. By using a dynamic approach, they map system calls to inputs for a deep learning model and utilize the explainable technique of Layer-wise Relevance Propagation to recognize which sequence parts are most significant in the decision-making process. By using a confusion matrix as a performance evaluation, they confirm that their method can swiftly and accurately identify malicious code.

Wang et al. \cite{wang2021exposing} focus on exposing vulnerabilities in malware detectors through explainability-guided evasion attacks that combine feature space manipulation with problem space obfuscation. They utilize a dataset of approximately 43,553 ELF binary files on Linux systems. Their research uses the model-agnostic explainability method SHAP to demonstrate how evasion attacks can be transferred from one detector to another. In another study, Mills et al. \cite{mills2019efficient} develop a lightweight malware detection system named NODENS, suitable for deployment on Raspberry Pi hardware. They test several ML-based algorithms on a Linux operating system, with the Random Forest algorithm performing optimally among them. This work utilizes a tree-based model to facilitate visual interpretation of the classification process, which enhances the end user's understanding of the output and aids in the individual development of the malware sample lifecycle. Due to the infrequent application of the Linux platform, there is limited research on explainable malware detection within the Linux domain. Studies addressing this topic are detailed in Table \ref{tab:Hardware/PDF/Linux}.

These Linux-oriented approaches show that explaining which code snippets or system calls are pivotal can help analysts better grasp the root causes of malicious behavior. However, the limited availability of Linux malware datasets—and the complexity of dynamic sandboxing—remains a bottleneck for broad adoption.

\subsection{Other Approaches}

\begin{table*}
    \centering
    \footnotesize
    \setlength{\tabcolsep}{0.8\tabcolsep}
    \def\arraystretch{1.5}
    \caption{Research Addressing Explainable Machine Learning in Malware Analysis.}
    \resizebox{\textwidth}{!}{%
    \begin{tabular}{|p{1.25cm}|p{3.5cm}|p{5.5cm}|p{3.5cm}|p{2.25cm}|}
    \hline
    \textbf{Paper/Year} & \textbf{Focus/Objective} & \textbf{Contribution} & \textbf{Limitation} & \textbf{XAI technique}\\
    \hline
    \multicolumn{5}{|l|}{\textbf{General Malware Approaches}}\\
    \hline
    Chen 2018~\cite{chen2018deep} & 
    Employ image-based static ML with LIME explanations &
    Visualizes malicious features in grayscale images for improved interpretability &
    Unverified against adversarial or obfuscated malware; lacks trust metrics &
    LIME \\
    \hline
    Briguglio and Saad. 2019~\cite{briguglio2019interpreting} & 
    Use LRP to interpret ML-based n-gram analysis in malware detection  &
    Compared LR, RF, NN using LRP for n-gram feature relevance &
    Lacks thorough interpretability framework analysis &
    Layer-wise relevance propagation (LRP) \\
    \hline
    Li et al. 2021~\cite{li2021novel} & 
    Propose LSH-based clustering for interpretable function-level malware classification &
    Groups similar assembly functions, offering built-in interpretability &
    Potential limitations not clearly discussed &
    Executable functions \\
    \hline
    Fidel et al. 2020~\cite{fidel2020explainability} & 
    Exploit SHAP “signatures” to detect adversarial vs. benign samples &
    Shows SHAP-value distributions differ for adversarial samples, enabling classifier-based detection &
    Method not easily transferable across detectors; SHAP’s non-differentiability is limiting &
    SHAP \\
    \hline
    Kumar and Subbiah 2022~\cite{kumar2022zero} & 
    Apply SHAP to identify top features in zero-day malware classification &
    Employs SHAP bar/waterfall plots to analyze false positives/negatives in zero-day detection &
    Does not deeply explore SHAP-based misclassification causes &
    SHAP \\
    \hline
    lee et al. 2022~\cite{lee2022automatic} & 
    Use reliability indicators + SHAP to refine alerts in large-scale security threats &
    Validated screening method with IDS/malware datasets, employing SHAP-driven feature insights &
    Omits comparison with alternate XAI tools &
    SHAP \\
    \hline
    Galli et al. 2024~\cite{galli2024explainability} & 
    Evaluate SHAP, LIME, LRP, attention to explain LSTM/GRU in behavioral malware detection &
    Proposed an XAI framework for behavioral models, assessing multiple explanation methods across diverse datasets &
    Unclear if results generalize beyond tested datasets &
    SHAP, LIME, LRP, and attention mechanisms \\
    \hline
    \end{tabular}}\\
    \vspace{1mm}
    \vspace{-4mm}
    \label{tab:other approaches}
\end{table*}

This section provides information regarding approaches that are not specific to any domain. Chen \cite{chen2018deep} leverages DL-based techniques for static malware classification to emphasize the importance of model transparency to gain user trust. They enhance their model's interpretability by utilizing LIME and adopting an image-based approach to visualize malware data. The study is conducted using three distinct datasets, where the model demonstrates high accuracy and a low false positive rate. For a practical demonstration of interpretability, the authors select an image from Lloyda.AA2 malware family and represent it with 200 super-pixel representations. They then identify which aspects of the malware images are crucial for the deep learning model's predictions. Their visual interpretation states that the red regions indicate the pixel regions that the model does not trust to contribute to the prediction. 

To evaluate interpretability techniques applied to ML-based malware detectors, Briguglio and Saad \cite{briguglio2019interpreting} explore how these techniques enhance N-gram analysis in the interpretation of machine-learning malware detectors. They focus on logistic regression, random forest, and neural network models, enhancing model confidence and feature significance. Specifically, they use the Layer-wise Relevance Propagation (LRP) technique to recognize the most important input nodes for classification.

Li et al. \cite{li2021novel} develop a novel ML-based model that classifies malware effectively and offers exceptional interpretability. They introduced a unique ML-based algorithm, the LSH-based clustering approach, which supports result visualization and interpretation that distinguishes it from other models in the field.

In their study on detecting evasion attacks, such as adversarial examples, Fidel et al. \cite{fidel2020explainability} utilize SHAP values as an innovative approach. They created SHAP signatures based on the premise that these signatures differ between benign and adversarial samples. Their findings confirm the initial hypothesis that variations in SHAP values in the classification model's final layer can effectively reveal the distribution of feature importance in classification outcomes. This method enhances the model's ability to identify adversarial examples, demonstrating a novel application of SHAP values in enhancing security measures.

Kumar and Subbiah \cite{kumar2022zero} conduct a static analysis using three different datasets to detect zero-day malware with ML-based algorithms. Among the algorithms tested, XGBoost achieves the highest accuracy and outperforms all other models. The authors utilize the SHAP bar and waterfall plots to identify the most significant features contributing to the model's predictions. They compare these top features across four categories of samples: False Positives (FP), False Negatives (FN), True Negatives (TN), and True Positives (TP). This comparison helps recognize misclassification categories, and the findings suggest that redistributing misclassified samples into their correct categories could significantly enhance the model's efficiency.

Lee et al. \cite{lee2022automatic} address large-scale threats to cybersecurity by leveraging IDS and malware datasets to validate the effectiveness of their proposed approach. Their method focuses on screening high-quality data to identify and rectify false predictions using reliability indicators. They incorporate the SHAP explainability technique to determine the contributions of individual features to specific outcomes. This approach identifies weaknesses in the existing AI models and enhances the detection of valuable alerts. By improving the accuracy of alert detection, the method allows human analysts to work more effectively and efficiently, which leads to prioritizing critical threats and optimizing response strategies.

Galli et al. \cite{galli2024explainability} address the critical need for transparency in AI systems used for malware detection. They develop and evaluate an XAI framework that applies to behavioral malware detection by employing DL models such as LSTM and GRU. These models analyze sequences of API calls to detect malicious activities. To make the models' decisions understandable and trustworthy, the paper investigates four different XAI techniques, i.e., SHAP, LIME, LRP, and attention mechanisms. The evaluation of these methods across three datasets (Mal-API-2019, API Call Sequences, and Alibaba Cloud Malware) shows their varying effectiveness in providing clear and useful explanations.

These generalized XAI solutions—covering n-gram analysis, signature-based detection, or adversarial defense—offer additional perspectives on explaining ML-driven malware detection. While each technique addresses a different niche (e.g., zero-day detection, adversarial resilience, multi-platform coverage), common XAI issues like scalability, consistency of explanations, and susceptibility to manipulation still arise.

In summary, explainable malware detection strategies span a diverse range of platforms and techniques—from gradient-based methods pinpointing critical file regions to model-agnostic tools offering broader coverage at the cost of local-only explanations to advanced image-based or rule-based systems. While these approaches strengthen trust, transparency, and analyst insight, they also face challenges in scaling to large datasets, handling obfuscated or adversarial samples, and balancing interpretability with accuracy. In the next section, we examine how these XAI methods can be enhanced against advanced threat tactics and integrated into practical cybersecurity frameworks.
\section{Future Research Directions}
\label{sec:futuredirections}

\begin{figure*}
    \centering
    \includegraphics[width=1\linewidth]{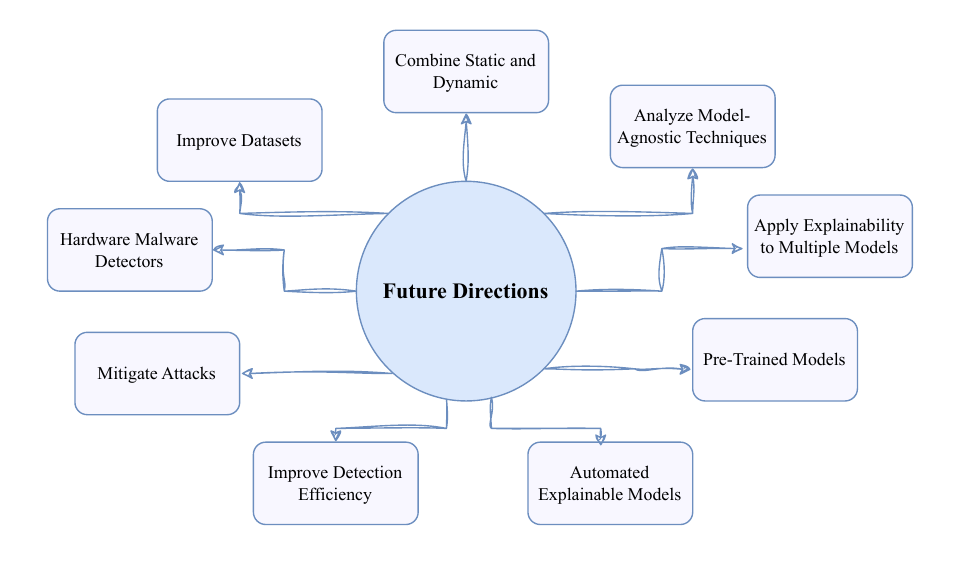}
    \caption{Future Directions for Explainable Malware Detection.}
    \label{fig:future}
\end{figure*}

Explainable ML is an evolving field with many ongoing challenges and opportunities for exploration. In the previous sections, we conducted an extensive review of various explainable ML techniques, with a particular focus on malware classification and detection. however, as the application of explainable methods in malware detection becomes increasingly prevalent, new challenges continue to emerge. As shown in Fig \ref{fig:future}, this section outlines several key challenges and potential research directions that researchers may pursue as future work in the area of explainable malware analysis.

\subsection{Improve Datasets}

Improving and updating malware datasets is a critical concern in the field of XAI. Many existing datasets are outdated and lack comprehensive coverage of current malware behaviors. These datasets often do not provide a sufficient volume of data for training XAI applications. For instance, previous research on explainable Android malware detection utilized the DREBIN dataset, which comprises 5,616 malicious instances and 121,329 benign instances \cite{arp2014drebin}. This imbalance, where benign instances significantly outnumber malicious ones, can hinder the training of effective models.

Moreover, the size of the current datasets is generally too small to train robust models. This field of research needs an unbiased, reasonably sized benchmark dataset that equally represents both benign and malicious behaviors. Accessing this kind of dataset is essential for evaluating explainable ML-based techniques and achieving reliable detection results. Furthermore, the DREBIN dataset, in particular, highlights the limitations of static analysis, pointing to the necessity for dynamic updates that support more comprehensive dynamic analyses. Additionally, there is potential for innovation in automated data generation and minimization techniques to accelerate the prediction process. For example, in hardware malware detection, researchers \cite{pan2021hardware} have generated trace data to facilitate hardware trace analysis and distinguish between malware and benign programs. Future work could focus on enhancing these techniques to streamline and speed up the predictive capabilities of malware detection systems.

\subsection{Combine Static and Dynamic}

This paper discussed explainability techniques in malware analysis, wherein researchers have primarily concentrated on static and dynamic analyses. Static analysis involves feature extraction and dimensionality reduction, processes that minimize information uncertainty and facilitate the analysis of malicious applications. Conversely, dynamic analysis focuses on training surrogate learning models. However, there is a notable absence of research on hybrid analysis, which combines elements of both to enhance the explainability of malware detection.

To address this gap, multiple ML-based classifiers will be leveraged to analyze both source code and runtime dynamic features. This dual approach aims to improve the efficiency and effectiveness of ML-based algorithms in distinguishing between benign and malicious applications. Moreover, to overcome the limitations in static and dynamic methodologies, researchers should also develop online and real-time explainable malware detection systems. These systems would continuously monitor the entire system to detect any possible malicious behavior or traces at any moment. Thus, the development of hybrid and online detection systems represents a significant research challenge in the field of explainable malware analysis.

\subsection{Analyze Model-Agnostic Techniques}

There is a need to explore various model-agnostic techniques that provide both local and global explanations, which can help develop fast-training and explainable models without sacrificing accuracy. One promising direction for future research is the automation of explainability when it is decoupled from the underlying machine learning model. This decoupling facilitates the easy replacement of both the explainable technique and the machine learning model itself.

The model-agnostic technique, LIME, is widely cited in research but comes with notable limitations. Its reliance on data sampling can lead to variability in explanations, making them potentially unstable and unreliable. Furthermore, if the local fidelity measure is inaccurate, the reliability of LIME’s explanations for distinguishing between malicious and benign samples is compromised. Additionally, LIME lacks guidance on the optimal number of features to use, which could affect the quality of its explanations.

To advance the field of explainable malware detection, paying attention to evaluating various model-agnostic techniques is essential. Future research could also focus on improving fidelity in explanations, which is crucial for maintaining reliability in rapidly evolving scenarios. Overall, model-agnostic methods represent a flexible and effective approach to enhancing malware detection through explainable ML.

\subsection{Apply Explainability Approach to Multiple Models}

Previous research in explainable ML within the malware detection domain has primarily focused on developing frameworks and applying specific explainability methods to those frameworks. However, there is a notable gap in the literature regarding the selection of explainability techniques for non-differentiable models. Theoretical findings suggest that under certain assumptions, various ML-based algorithms can yield similar decision functions. This similarity raises a critical question: how does one select the most appropriate explainable technique for a given malware detection process?

Hence, conducting thorough analyses and evaluations of how different explainability techniques influence the explanations generated by a specific framework. Such research can demonstrate that the chosen explainability technique fits the model and outperforms alternative methods in clarity and effectiveness. Enhancing the understanding of the applicability and efficiency of various explainable methods in malware detection leads to more robust and transparent systems.

\subsection{Use Pre-Trained Models}

While neural networks are powerful tools for modeling, their black-box nature makes them difficult to interpret, which poses a significant challenge in fields such as malware detection. In the research regarding this, the authors \cite{pan2020hardware} have developed a framework that utilizes neural networks to facilitate interpretable malware detection, which has innovative approaches to this issue.

Looking forward, in the malware detection domain, it would be advantageous to leverage existing explainable pre-trained models rather than building new models from scratch. This approach can save considerable time that would otherwise be spent collecting data and training models and enhances the efficiency of detecting malicious activities in systems that may already be compromised. 

In other words,  utilizing pre-trained models can accelerate the deployment of malware detection systems and improve their effectiveness by integrating advanced, pre-learned features into the detection process. At present, there is no well-established, publicly available pre-trained explainable model that is widely adopted specifically for malware detection. While the concept of pre-trained” and “explainable models exists in other domains (e.g., NLP, computer vision), the malware analysis community has not converged on a standard, large-scale pre-trained model that includes built-in explainability for detecting malicious code or behavior.

Building a comprehensive pre-trained explainable model for malware detection is hampered by challenges such as limited access to large, high-quality datasets due to privacy and proprietary concerns, as well as the constantly evolving nature of malware which requires frequent retraining and adaptation. Additionally, integrating explainability mechanisms at scale introduces complexity and may reduce performance if not carefully designed and maintained.

\subsection{Automated Explainable Models}

One promising direction for future research in the field of malware detection involves implementing more automated explainability models. The goal is to enhance user trust in black-box models, such as those based on DL techniques, which are currently not automatically interpretable. Most existing research focuses on interpreting the results of malware detection after the detection has already occurred. This method leaves a gap in real-time understanding and response, which automated explainability aims to fill.

Moreover, achieving an optimal balance between accuracy and explainability continues to be a significant challenge. Automated explainability could help bridge this gap by providing insights into the decision-making process of complex models in real time. Additionally, there is a clear need for more research focused on quantitative-level evaluation of these explainable models. Such evaluations would assess the interpretability and how the introduction of explainability affects the overall performance of the detection system. 

\subsection{Improve Detection Efficiency}

A valuable future direction in explainable malware detection is to enhance the design methodologies of malware detectors so that the explanations they generate can assist professionals in more accurately characterizing malware attacks. For example, wang et al.  \cite{wang2016trafficav} involve extracting features and employing a decision tree to develop a model capable of determining the maliciousness of applications.

Looking forward, the implementation of pruning strategies in decision trees presents a promising avenue for enhancing the efficiency of these detection models. Pruning optimizes the tree structure by removing superfluous or minimally informative branches, thereby simplifying the model. This optimization can accelerate the processing time and enhance the accuracy by focusing the model’s analysis on the most significant features. 

\subsection{Mitigate attacks}

In recent research, the primary focus has been on black-box attacks, gradient-based attacks, evasion attacks, and poisoning attacks. Evasion attacks involve manipulating malicious input samples during the training phase to circumvent detection by a trained system, and it requires access to the model. Poisoning attacks compromise the integrity of training data by introducing incorrect data since it can mislead the learning process of ML models. This corruption of training data severely undermines the entire training process.

In both gradient-based and poisoning attacks, it is assumed that the attacker has knowledge of the feature space used by the target. Future research in the field of explainable ML should explore defense mechanisms against these types of attacks and develop generic mitigation methods. Moreover, while current attacks typically use either static or dynamic approaches, future attacks might utilize a hybrid approach that integrates both strategies. As malware data continuously evolves, implementing attacks in online detection systems could pose significant challenges for attackers trying to intercept or manipulate high-speed continuous data compared to data stored on devices.

The study by Scalas et al.  \cite{scalas2021malware} highlights the use of system API calls as effective features for detecting attack strategies. Future research could assess the susceptibility of system API calls to attacks and explore whether this detection strategy limits the number of features that attackers can feasibly manipulate. 

\subsection{Hardware Malware Detectors}

Research on explainable hardware-based malware detection is currently limited, which presents significant opportunities for future investigation. One potential avenue for advancement involves the design of efficient and explainable hardware malware detectors. These systems could automate the trace selection process and reduce prediction time while maintaining high accuracy in differentiating between malware and benign programs.

Another area for exploration is the development of debugging architectures that enhance malware detection capabilities. This could include the design of embedded trace buffers and the utilization of hardware performance counters. These tools would help identify the most informative traces for use in explainable machine learning applications within the malware detection field. This focus can enhance the efficiency and effectiveness of malware detection systems and make them more accessible and interpretable for cybersecurity professionals.

\section{Conclusion}
ML-based techniques play a crucial role in cybersecurity, yet these data-driven frameworks are susceptible to exploitation, misdirection, and circumvention. Explainability is essential to enhance the transparency of these models and build trust in order to deploy them effectively for malware analysis.

This paper discusses explainable ML in malware analysis and reviews state-of-the-art approaches. We provide an in-depth examination of explainable malware classification and detection methods, summarizing the work of researchers to date. Our study systematically organizes various explainable malware-based approaches, making this information more accessible to researchers and others interested in this field.

We conclude the survey by identifying open research challenges and future directions in explainable malware analysis. This survey serves as a comprehensive guide for researchers exploring explainable malware detection, offering insights into the current landscape and stimulating research in unexplored domains within this dynamic and evolving field.

\bibliographystyle{IEEEtran}
\bibliography{references}

\begin{IEEEbiography}[{\includegraphics[width=1.1in,height=1.5in,clip,keepaspectratio]{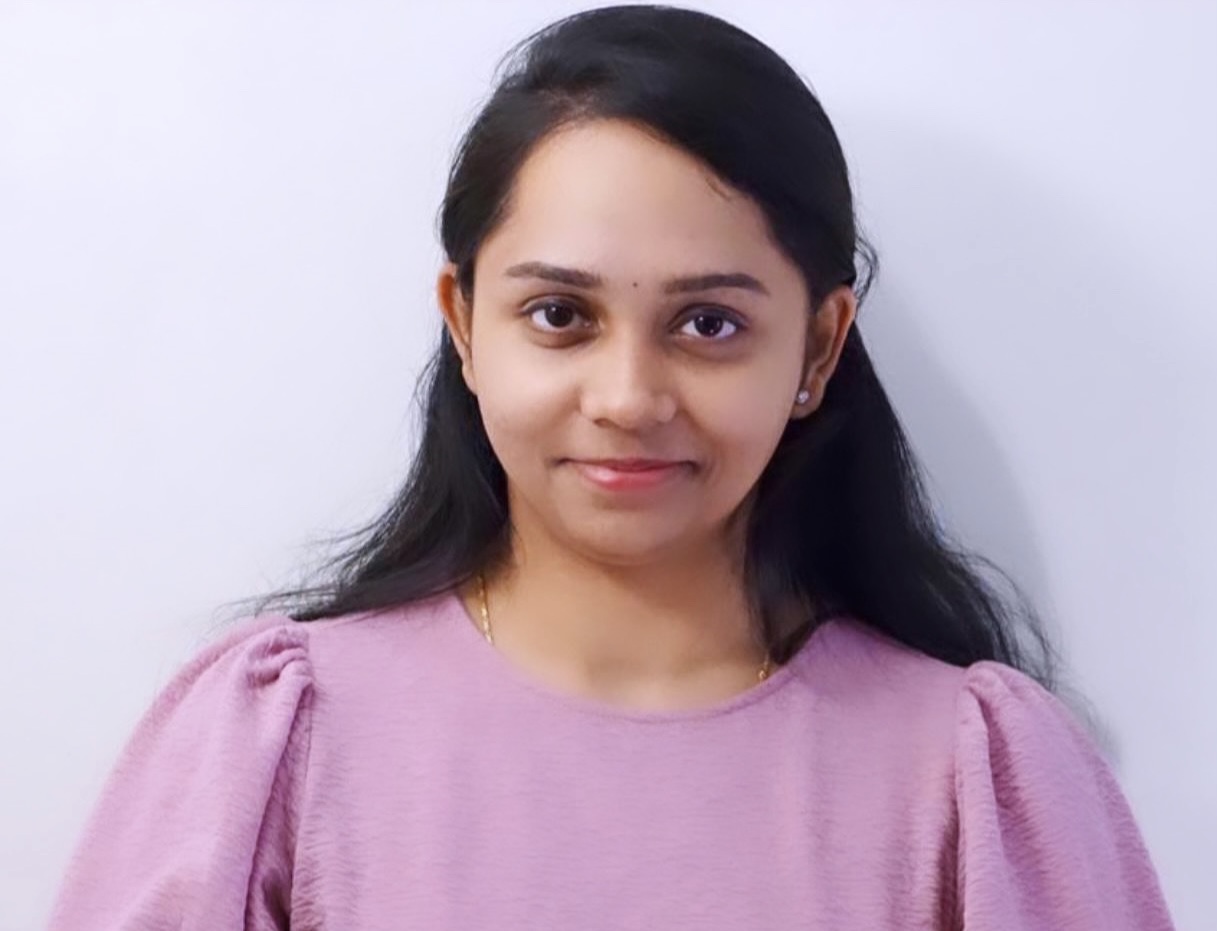}}]{Harikha Manthena}
received her B.Tech degree in Computer Science from Andhra University in 2016. She worked in Fidelity Information Services (FIS) as a software engineer in Bangalore, India. She received her M.S. in Computer Science from North Carolina A\&T State University in 2022. Her research interest includes explainable/interpretable machine learning-based malware analysis in the cloud. 
\end{IEEEbiography}

\begin{IEEEbiography}[{\includegraphics[width=1.1in,height=1.5in,clip,keepaspectratio]{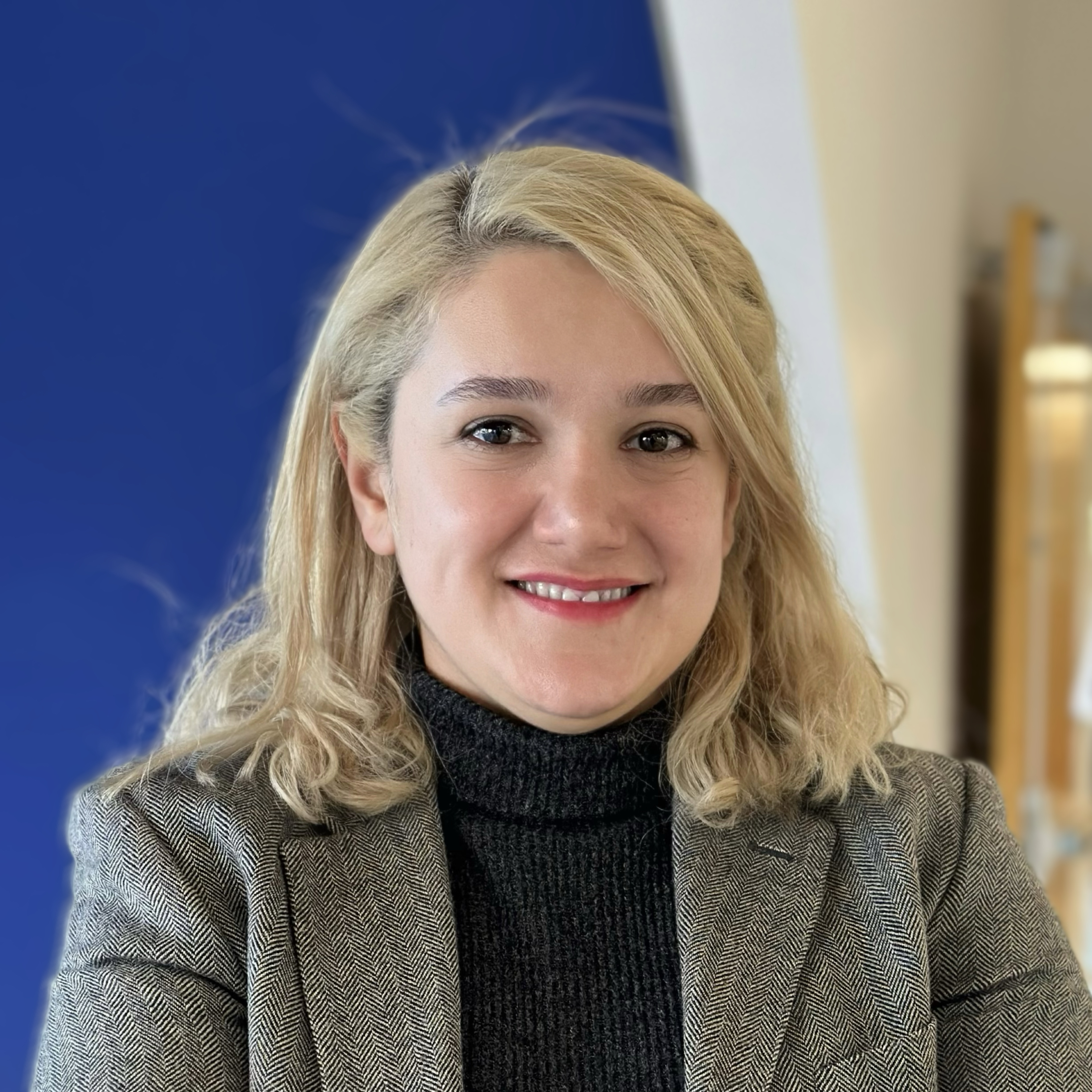}}]{Shaghayegh Shajarian} received her B.S. and M.S. degrees in Computer Software Engineering from the University of Mazandaran, Iran, and the Science and Research Branch, Azad University, Tehran, Iran, in 2016 and 2019, respectively. She is currently pursuing her Ph.D. degree in Computer Science at North Carolina A\&T State University. Her research interests include autonomous networks, network management, and applied AI/ML.
\end{IEEEbiography}

\begin{IEEEbiography}[{\includegraphics[width=1.1in,height=1.5in,clip,keepaspectratio]{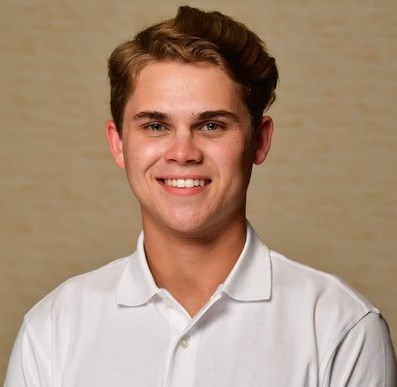}}]{Jeffrey C Kimmell}
graduated with a M.S. and B.S. in Computer Science from Tennessee Tech University in 2022 and 2021, respectively. His research interest includes deep learning and AI-based malware analysis in the cloud and its explainability aspects. He currently works at Oak Ridge National Laboratory. 
\end{IEEEbiography}

\begin{IEEEbiography}[{\includegraphics[width=1.1in,height=1.5in,clip,keepaspectratio]{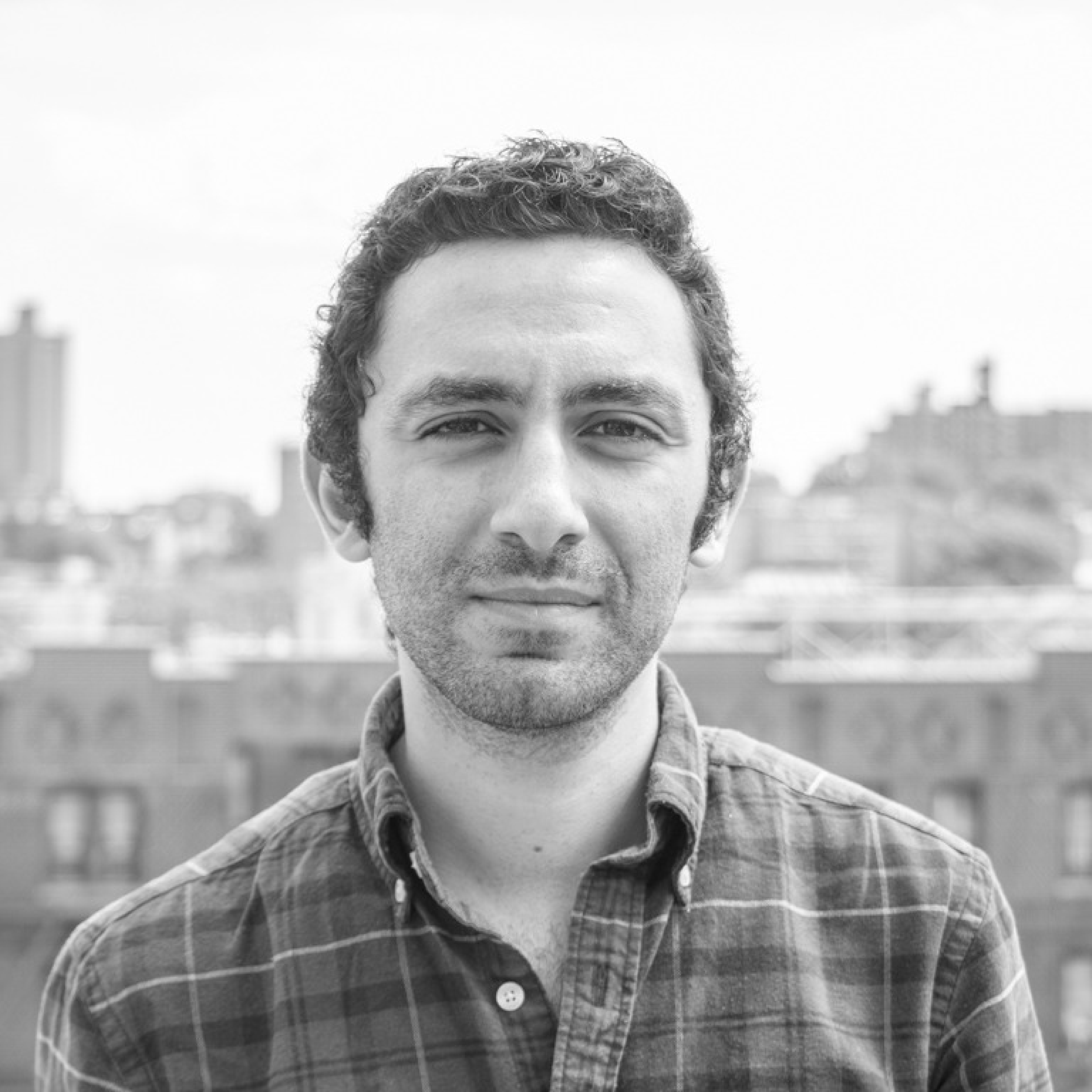}}]{Mahmoud Abdelsalam}
received the M.Sc. and Ph.D. degrees from the University of Texas at San Antonio (UTSA), in 2017 and 2018, respectively. He was working as a Postdoctoral Research Fellow with the Institute for Cyber Security (ICS), UTSA, and as an Assistant Professor with the Department of Computer Science, Manhattan College. He is currently working as an Assistant Professor with the Department of Computer Science, North Carolina A\&T State University. His research interests include computer systems security, anomaly and malware detection, cloud computing security and monitoring, cyber-physical systems security, and applied ML.
\end{IEEEbiography}

\begin{IEEEbiography}[{\includegraphics[width=1.1in,height=1.5in,clip,keepaspectratio]{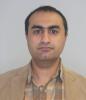}}]{Sajad Khorsandroo} received his Ph.D. in Computer Science from the University of Texas at San Antonio in 2019. He is currently an Assistant Professor in the Department of Computer Science at North Carolina A\&T State University. His research spans systems, cybersecurity, and applied AI/ML, supported by funding from federal and state agencies as well as industry collaborators, including the National Science Foundation (NSF), Department of Defense (DoD), Carolina Cyber Network, and Palo Alto Networks, Inc.
\end{IEEEbiography}

\begin{IEEEbiography}[{\includegraphics[width=1.1in,height=1.5in,clip,keepaspectratio]{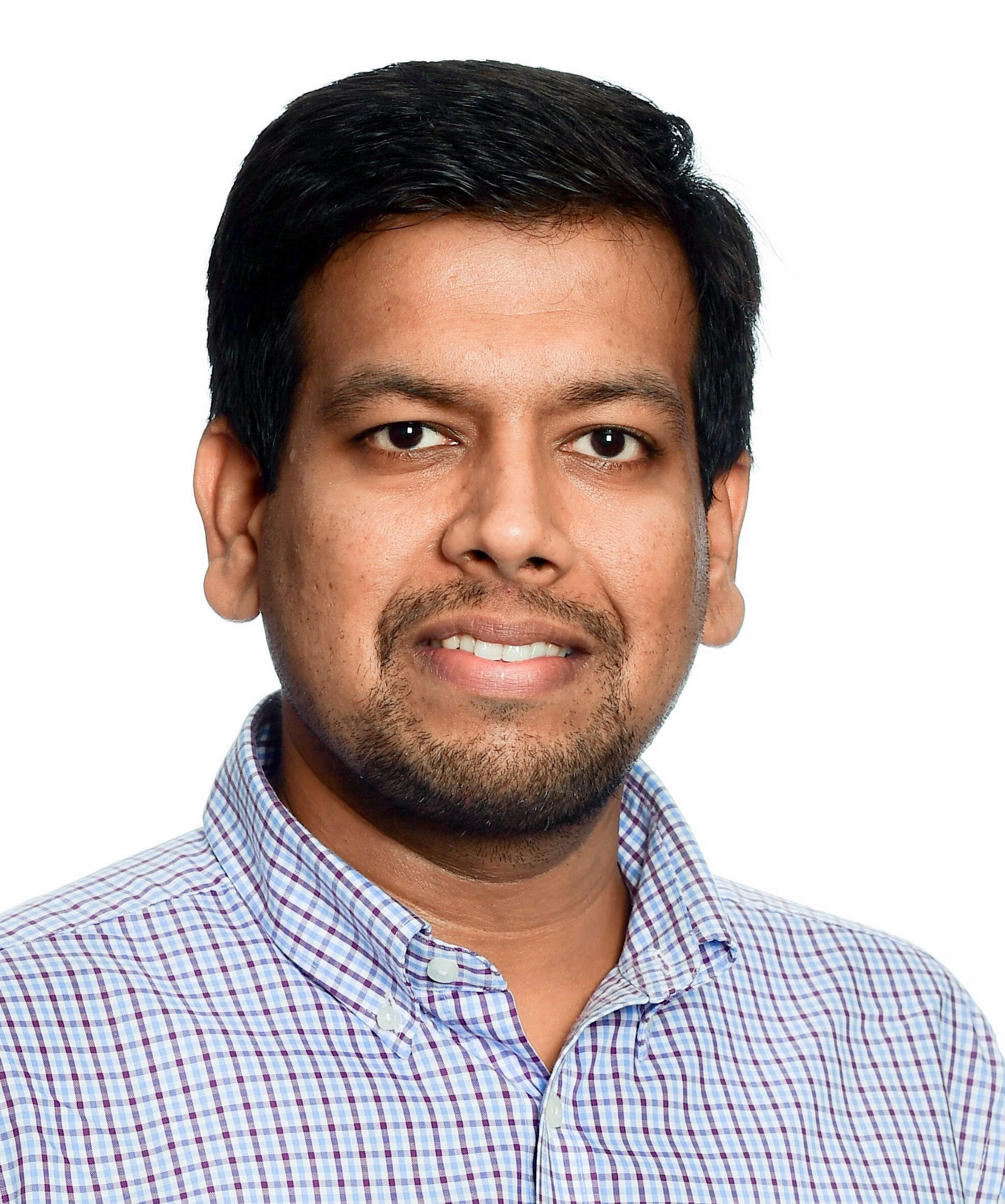}}]{Maanak Gupta} (Senior Member, IEEE) is an Associate Professor in Computer Science at Tennessee Technological University, Cookeville, USA. He received his M.S. and Ph.D. in Computer Science from the University of Texas at San Antonio (UTSA) and has also worked as a postdoctoral fellow at the Institute for Cyber Security (ICS) at UTSA. His primary area of research includes security and privacy in cyberspace, focused on studying foundational aspects of access control, malware analysis, AI and machine learning-assisted cyber security, and their applications in technologies, including cyber-physical systems, cloud computing, IoT, and Big Data. His research has been funded by the US National Science Foundation (NSF), NASA, and the US Department of Defense (DoD), among others.
\end{IEEEbiography}

\EOD

\end{document}